\documentclass[
reprint,
superscriptaddress,
nofootinbib,
aps, pra
]{revtex4-1}
\usepackage{subcaption}
\usepackage{float}
\usepackage{amsmath, amsfonts}
\usepackage{MnSymbol}
\usepackage{bm}	
\usepackage{soul}
\usepackage{comment}
\usepackage{bbm}				 
\usepackage{braket}				 
\usepackage{mathtools}			 
\usepackage{graphicx} 			 
\usepackage{pst-plot}			 
\usepackage[hidelinks]{hyperref} 
\usepackage{color}
\usepackage{CJKutf8}
\newcommand{\nn}{\nonumber}	

\DeclareMathOperator{\sinc}{sinc}

\begin{document}

\title{Cavity mode dephasing via the optomechanical interaction with an acoustic environment}

\author{Qidong Xu}
\email[]{qidong.xu.gr@dartmouth.edu}

\author{M. P. Blencowe}
\email[]{miles.p.blencowe@dartmouth.edu}

\address{Department of Physics and Astronomy, Dartmouth College, Hanover, New Hampshire 03755, USA}

\date{\today}

\begin{abstract}
We consider an optomechanical system comprising a single cavity mode and a dense spectrum of acoustic modes and solve for the quantum dynamics of initial cavity mode Fock (i.e., photon number) superposition states and thermal acoustic states. The optomechanical interaction results in dephasing without damping and bears some analogy to gravitational decoherence.  For a cavity mode locally coupled to a one-dimensional (1D) elastic string-like environment or two-dimensional (2D) elastic membrane-like environment, we find that the dephasing dynamics depends respectively on the string length and membrane area--a consequence of an infrared divergence in the limit of an infinite-sized string or membrane. On the other hand, for a cavity mode locally coupled to a three-dimensional (3D) bulk elastic solid,  the dephasing dynamics is independent of the solid volume (i.e., is infrared finite), but dependent on the local geometry of the coupled cavity--a consequence of an ultraviolet divergence in the limit of a ``pointlike" coupled cavity. We consider as possible respective  realizations for the cavity-coupled-1D and 2D acoustic environments, an LC oscillator capacitively coupled to a partially metallized strip and a cavity light mode interacting via light pressure with a membrane.          
\end{abstract}

\maketitle

\section{Introduction}
\label{Introduction}
Cavity optomechanical systems have received considerable attention over the past decades, with applications ranging from the detection of classical gravity waves in the macroscopic domain to the generation and detection of  quantum states of mechanical oscillators in the nano-to-mesoscale regimes~\cite{aspelmeyer2014,bowen2015}. Most investigations deliberately consider one or at most a few cavity modes interacting similarly with one or at most a few mechanical modes, with a notable exception involving the consideration of interacting optical and acoustic waves coexisting in bulk, crystalline solids~\cite{renninger2018}.

In this present work, we shall take as our starting point the following  Hamiltonian:
\begin{align}
H  =& \hbar \Omega \left(a^\dag a +\frac{1}{2}\right) \left(1+ \sum_i \lambda_i \left(b_i +b_i^\dag\right)  \right) \nn \\
+& \sum^N_{i=1} \hbar \omega_i \left(b_i^\dag b_i +\frac{1}{2} \right),
\label{Hamiltonian}
\end{align}
where here $a,\, a^{\dag}$ are the annihilation/creation operators for a cavity mode with frequency $\Omega$, while the $b_i,\, b_i^{\dag}$ are the annihilation/creation operators for $N$ mechanical modes. The cavity and mechanical modes are coupled via the standard optomechanical interaction with coupling constant parameters $\hbar\Omega\lambda_i$. Our particular focus will be on the effective dynamics of the {\it single} cavity mode system interacting with {\it many} (i.e., $N\ggg 1$) mechanical modes, with the latter viewed as an acoustic, environmental bath for the cavity system. In contrast to the usual quantum Brownian motion model, where the system-bath coupling is bilinear in their respective creation/annihilation coordinates, Hamiltonian (\ref{Hamiltonian}) does not result in energy damping of the cavity mode system. This is a consequence of the fact that the system Hamiltonian commutes with the interaction Hamiltonian term. On the other hand, dephasing does result for initial superpositions of energy eigenstates of the cavity system; for this reason, Ref. 
\cite{gardiner2004} terms Eq. (\ref{Hamiltonian}) the ``phase damped oscillator", and provides an approximate solution to the cavity system reduced density matrix dynamics via a master equation approach.

As we shall show, the effective dynamics for cavity system reduced density matrix can in fact be solved  exactly up to a summation over bath modes, while the latter summation can be carried out approximately for certain bath spectral densities; the method of solution is based on that of Refs. \cite{bose1997,bose1999scheme}, which consider a single cavity mode interacting with a single mechanical mode, and which again utilizes the fact that the system and interaction term Hamiltonians commute.   

Our interest in the Hamiltonian (\ref{Hamiltonian}) and the resulting dephasing dynamics of the cavity mode system reduced state stems from its analogue connection with gravitationally induced decoherence \cite{blencowe2013,xu2020}. In the weak gravitational field regime, the leading order term in the interaction action involving  a scalar matter field $\phi(x)$ system and gravitational metric deviation $h_{\mu\nu}$ from Minkowski space environment takes the form
\begin{equation}
 S_{{I}}=\sqrt{8\pi G}\int d^4 x T^{\mu\nu}(\phi)h_{\mu\nu}
 \label{gravinteq}
\end{equation}
 in natural units $\hbar=c=1$, where $T^{\mu\nu}(\phi)$ is the scalar field energy-momentum tensor. This interaction term can result in the dephasing of scalar field energy superposition states without energy damping \cite{blencowe2013,anastopoulos2013}, just as for the cavity mode quantum dynamics following from Hamiltonian (\ref{Hamiltonian}) \cite{xu2020}.

However, the cavity system dynamics following from the Hamiltonian (\ref{Hamiltonian}) interpreted as modeling cavity optomechanical bath systems is of interest in its own right, particularly the consequences  of the acoustic environment spatial dimension and size for the cavity mode energy quantum superposition dephasing dynamics. We shall find that for 1D and 2D elastic ``string" and ``membrane" acoustic environments respectively, the cavity system dephasing dynamics depends on the geometric size of the environment--a consequence of an infrared (IR) divergence in the limit as the environment size tends to infinity. In contrast, for a bulk, elastic 3D acoustic environment (which shares the same Ohmic spectral density as for the gravitational wave environment \cite{blencowe2013}), the cavity dephasing dynamics depends on the size of the optical cavity system embedded within the 3D elastic medium--a consequence of an  ultraviolet (UV) divergence in the limit as the size of the cavity tends to zero, i.e., becomes pointlike. 

Infrared divergences arising from long wavelength acoustic flexural modes of membrane-like structures in the infinite size limit are also encountered in other contexts, for example the thermal expansion of 2D crystals \cite{michel2015theory} and atom--membrane surface interactions \cite{clougherty2014quantum,sengupta2016infrared,clougherty2017infrared,clougherty2017infraredtemp,sengupta2017radiative,sengupta2019theory}.
 
 In Sec. \ref{dephasingsec}, we solve for the cavity system reduced density matrix evolution following from the time dependent Schr\"{o}dinger equation with Hamiltonian (\ref{Hamiltonian}) in the Fock state (i.e., photon number) basis for both ohmic ($s=1$) and subohmic ($s=0, -1$) bath spectral densities [see Eq. (\ref{generalspectraldensity})], and with the oscillator environment in an initial thermal state. This section extends the analysis of Ref. \cite{xu2020}, which considers only the Ohmic case and infinite-sized environment.  In Sec. \ref{1dcircuit}, we consider a model cavity-acoustic environment optomechanical system realization involving an LC oscillator capacitively coupled to a partially metallized, long elastic strip and show how this system maps onto the subohmic $s=-1$ case. Section \ref{2dmembrane} considers another model system consisting of an optical cavity interacting via light pressure with a large, square elastic membrane \cite{thompson2008}, which maps onto the subohmic $s=0$ case; both Secs \ref{1dcircuit} and \ref{2dmembrane} explore quantitatively by considering example, experimentally feasible device parameter values, the cavity mode quantum dephasing dynamics dependence on the acoustic environment size, i.e., the elastic strip length and side dimension of the square membrane. Sec. \ref{conclusionsec} gives a concluding discussion.

\section{Cavity Dephasing Dynamics}
\label{dephasingsec}
Our starting point is the standard single cavity  mode optomechanical Hamiltonian (\ref{Hamiltonian}), but with a bath of mechanical oscillator modes labelled by the index $i=0,1, 2,\dots,N\ggg 1$, instead of the usually considered single mode case  \cite{aspelmeyer2014}. Hamiltonian (\ref{Hamiltonian}) neglects cavity-mechanical oscillator bath interaction terms of the form $a^2 (b_i+b_i^{\dag})$ and $a^{\dag 2} (b_i+b_i^{\dag})$, which describe for example two photons annihilating and creating a bath phonon ($a^2 b_i^{\dag}$), or conversely a bath phonon annihilating and creating two cavity photons ($a^{\dag 2} b_i$). As we shall see later below in Secs. \ref{1dcircuit} and \ref{2dmembrane}, such terms can be neglected since the coupling constant $\lambda_i$ is suppressed for phonon wavelengths much smaller than the cavity size.   

We now briefly review the steps for solving the time-dependent Schr\"{o}dinger equation with Hamiltonian (\ref{Hamiltonian}) \cite{bose1997,bose1999scheme,xu2020}. We assume that the cavity mode system can be prepared in an initial product state with the bath, the latter of which is assumed to be in a thermal state: $\rho_{\mathrm{initial}}=\rho_{c} \otimes \rho_{\mathrm{bath}}$. The cavity system initial state is decomposed in terms of the Fock (i.e., number) state basis, $\rho_c=\sum_{n,n'}c_{n n'}|n\rangle\langle n'|$, and the thermal bath state expressed in a coherent state basis:
\begin{align}
\rho_{\mathrm{bath}} =& \prod_ i \frac{1}{\pi\left(e^{\beta \hbar \omega_i} - 1\right)} \int d \alpha_i^2 \exp\Big(-|\alpha_i|^2 \nn \\
\times& \left(e^{\beta \hbar \omega_i}-1\right) \Big)  |\alpha_i\rangle \langle \alpha_i|,
\label{rhothermeq}
\end{align}
where $\beta^{-1}=k_BT$, with $k_B$ Boltzmann's constant and $T$ the bath temperature. Solving first the Schr\"{o}dinger equation for an initial basis state $|n,\{\alpha_i\}\rangle$ and then tracing out the bath, we obtain for the reduced state of the cavity mode: $\rho_c(t)=\sum_{n,n'}c_{n n'}|n(t)\rangle\langle n'(t)|$, where the time-dependent outer product is
\begin{align}
&|n(t)\rangle \langle n'(t)| = |n\rangle \langle n'|\cr &\times\exp\Bigg(-it \left[\Omega (n-n') 
- (n+n'+1)(n-n')\sum_i \frac{\left(\Omega \lambda_i\right)^2}{\omega_i}\right]\cr
&- i(n+n'+1)(n-n')\sum_i \left(\frac{\Omega\lambda_i}{\omega_i}\right)^2\sin( \omega_i t)\cr
&-2(n-n')^2\sum_i \left(\frac{\Omega \lambda_i}{\omega_i} \right)^2 \coth\left( \frac{\beta \hbar \omega_i}{2}\right)\sin^2\left(\frac{\omega_i t}{2} \right)
\Bigg).
\label{outerprodeq}
\end{align}
Note that this outer product is time-independent for $n=n'$, a consequence of the fact that the system oscillator Hamiltonian commutes with the system-bath interaction Hamiltonian.

We now discuss the various terms appearing in Eq. (\ref{outerprodeq}).  The first imaginary term $-i\Omega (n-n')t$ in the argument of the exponential is just the free cavity oscillator system evolution. The second imaginary term gives rise to a cavity frequency renormalization  $\Omega'=\Omega-\sum_i(\Omega\lambda_i)^2/\omega_i$ [from the $(n-n')$ part],  as well as an induced 
Kerr nonlinear self-interaction [from the $\left(n^2-n'^2\right)$ part]  in the oscillator Hamiltonian: 
\begin{align}
H=\hbar \Omega a^\dag a + \hbar\Lambda_{\mathrm{kerr}} (a^\dag a)^2,
\label{kerroscillator}
\end{align}
where $\Lambda_{\mathrm{kerr}}=-\sum_i(\Omega\lambda_i)^2/\omega_i$. The third imaginary term cancels the just-described second imaginary term in the short time limit $t\rightarrow 0$, while it decays to zero as $t$ increases due to the oscillating sine term; later below, we give a more quantitative specification of the short and long time regimes.
Finally, the fourth, real term in the argument of the exponential in Eq. (\ref{outerprodeq}) can result in dephasing, causing the off-diagonal terms of the system reduced density  operator in the number state basis to decrease with increasing time. 
 
In order to obtain a more quantitative understanding of the time dependent behavior of the various terms appearing in the outer product expression (\ref{outerprodeq}), we shall now approximate the discrete sum over the acoustic bath modes with a continuous frequency integral. This necessarily requires $N\ggg 1$ for a sufficiently dense bath frequency spectrum. We shall assume the following bath spectral density approximation with exponential cut-off set by some upper frequency scale $\omega_u$:
\begin{align}
\pi\sum_i \lambda_i^2 f(\omega_i)\delta(\omega-\omega_i)\approx C\omega^s f(\omega)e^{-{\omega}/{\omega_u}},
\label{generalspectraldensity}
\end{align}
where the function $f(\omega)$ is determined by the $\omega_i$ dependence of a given term in argument of the exponential in Eq. (\ref{outerprodeq}), and $C$ is a frequency-independent coupling strength constant. Following common convention \cite{leggett}, we term optomechanical cavity-acoustic bath systems with exponent $s=1$ ``ohmic" and systems with exponent $s<1$ ``subohmic".  Depending on the value of the exponent $s$ and the form of $f(\omega)$, an upper cut-off may be required in order to regularize a possible UV divergence as $\omega\rightarrow\infty$. For the concrete example optomechanical model realizations in Secs. \ref{1dcircuit} and \ref{2dmembrane},  we will see that an upper cut-off arises naturally through a suppression of the optical mode system-acoustic bath coupling when the acoustic phonon wavelength becomes smaller than a characteristic optical cavity system dimension. Note that the functional form of the upper cut-off dependencies for these  concrete examples is not in fact of the same exponential form as assumed in Eq. (\ref{generalspectraldensity}). Nevertheless, it is still informative to consider the commonly-used exponential cut-off since it readily allows closed form analytical expressions for the various summation terms appearing in Eq. (\ref{outerprodeq}) approximated as integrals. 

Furthermore, a lower frequency cut-off, which we denote as $\omega_1\, (\ll\omega_u)$, may be required depending on the value of the exponent $s$ and form of the function $f(\omega)$, in order to regularize a possible IR divergence as $\omega\rightarrow 0$. For the model realizations considered in the following sections, a lower frequency cut-off arises naturally as the fundamental, lowest frequency mode $\omega_1$ of the acoustic environment medium which has a finite size. 

Using the spectral density approximation Eq.~(\ref{generalspectraldensity}), the two imaginary, induced phase terms in Eq. (\ref{outerprodeq}) can be evaluated approximately analytically by expressing them in terms of the incomplete Gamma function $\Gamma(s,z)=\int_z^{\infty} dx x^{s-1}e^{-x} $:
\begin{align}
&it(n+n'+1)(n-n')\sum_i \frac{\Omega^2 \lambda_i^2}{\omega_i}\nn\\
&\approx it (n+n'+1)(n-n')\frac{C\Omega^2}{\pi} \int_{\omega_1}^{\infty} d\omega \omega^{s-1}e^{-{\omega}/{\omega_u}}  \nn \\
&= it(n+n'+1)(n-n') \frac{C \Omega^2 \omega_u^s}{\pi} \Gamma\left(s, \frac{\omega_1}{\omega_u}\right), 
\label{analyticalkerr}
\end{align}
and
\begin{align}
-&i(n+n'+1)(n-n') \sum_i \frac{\Omega^2 \lambda_i^2}{\omega_i^2} \sin(\omega_i t) \nn \\
\approx& -i(n+n'+1)(n-n')\frac{C\Omega^2}{\pi} \int_{\omega_1}^{\infty} d\omega \omega^{s-2} \sin(\omega t) e^{-{\omega}/{\omega_u}} \nn \\
=&-i(n+n'+1)(n-n') \frac{C\Omega^2 \omega_u^{s-1}}{\pi} \nn \\
&\times {\mathrm{Im}}\left[(1-i\omega_u t)^{1-s} \Gamma\left(s-1, \frac{\omega_1}{\omega_u}(1-i\omega_u t) \right) \right].
\label{analyticalsecphase}
\end{align}

The real, induced dephasing term in Eq. (\ref{outerprodeq}), with the spectral density Eq.~(\ref{generalspectraldensity}), can only be approximated analytically in certain time range limits; we will consider the high temperature limit defined as $k_B T\gg \hbar/t$ (equivalently $t\gg \beta\hbar$), for which  the $\coth$ function can be expanded to leading order. The dephasing term can then similarly be expressed approximately in terms of incomplete Gamma functions:
\begin{align}
&-2 (n - n')^2 \sum_i \left( \frac{\Omega \lambda_i}{\omega_i}\right)^2 \coth\left( \frac{\beta \hbar \omega_i}{2}\right)  \sin\left(\frac{\omega_i t}{2}\right)^2\nn \\
\approx&-\frac{2C \Omega^2}{\pi} (n - n')^2 \int_{\omega_1}^{\infty} d\omega \omega^{s-2} \coth\left( \frac{\beta \hbar \omega}{2}\right)\nn\\
&\times\sin\left( \frac{\omega t}{2}\right)^2 e^{-{\omega}/{\omega_u}} \nn \\
\approx& -\frac{2C \Omega^2}{\pi} (n - n')^2 \int_{\omega_1}^{\infty} d\omega \omega^{s-2} \frac{2}{\beta \hbar \omega} \sin\left( \frac{\omega t}{2}\right)^2 e^{-{\omega}/{\omega_u}} \nn \\
=& -\frac{2C \Omega^2}{\pi} (n - n')^2 \frac{\omega_u^{s-2}}{\beta \hbar} \Bigg\{\Gamma\left( s-2, \frac{\omega_1}{\omega_u}\right) \nn \\
&- {\mathrm{Re}}\bigg[(1-i\omega_u t)^{2-s}  \Gamma\left(s-2, \frac{\omega_1}{\omega_u} (1- i\omega_u t) \right) \bigg] \Bigg\}.
\label{analyticaldephasing}
\end{align}

\begin{figure}
\begin{center}
\includegraphics[width=2.5in]{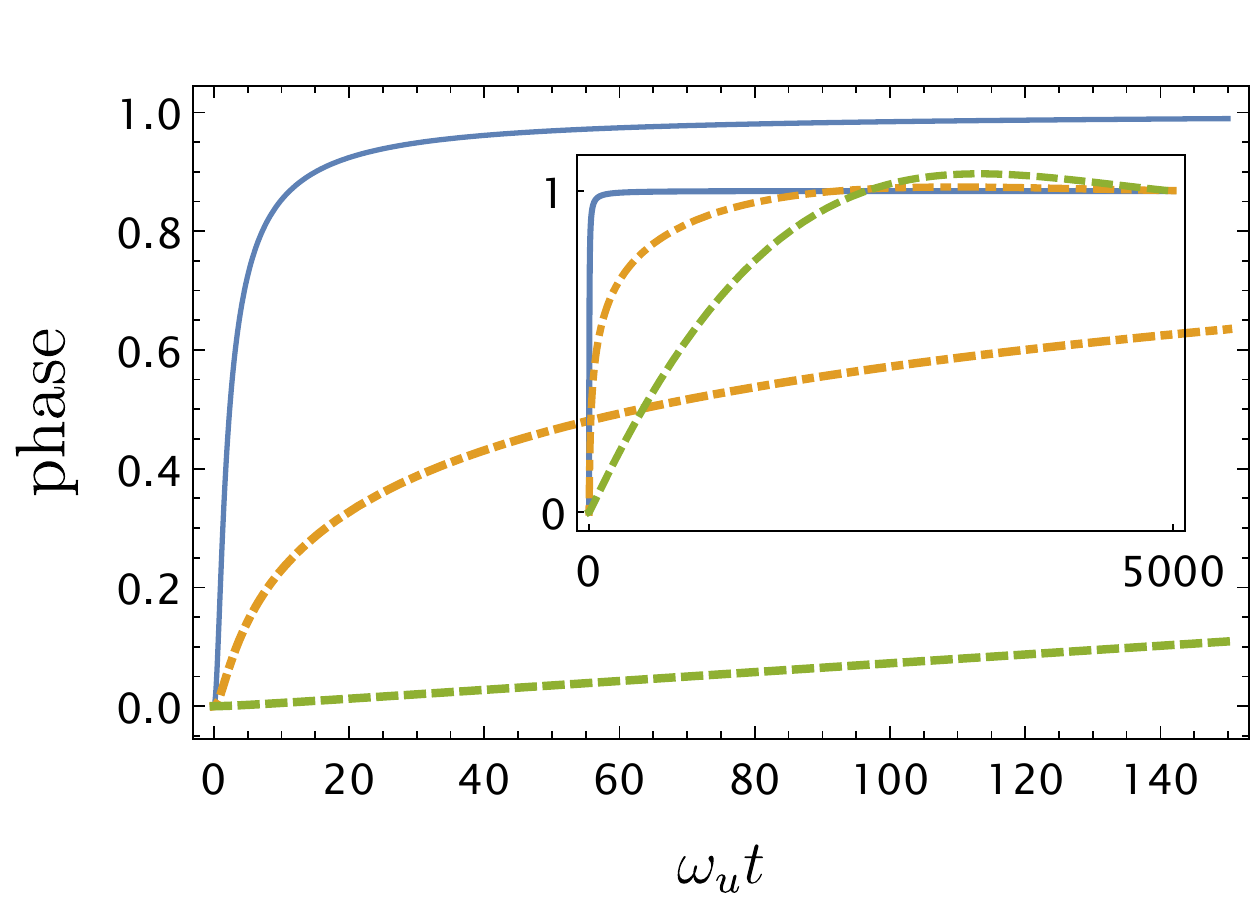} 
\includegraphics[width=0.8in]{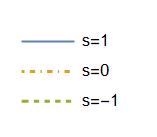}
\caption{\label{inducedphasefig}Sum of the two induced phase terms Eq.~(\ref{analyticalkerr}) and Eq.~(\ref{analyticalsecphase}) divided by its long time ($t\gg \omega_1^{-1}$) analytical expression as a function of dimensionless time $\omega_u t$, where we set $\omega_1/\omega_u = 0.001$. The inset gives the same normalized phase terms plotted over much longer timescales, indicating the expected approach to 1, hence validating the analytical approximation in the long time limit.}
\end{center}
\end{figure}

In the following three subsections, we shall explore the time dependencies of Eqs. (\ref{analyticalsecphase}) and (\ref{analyticaldephasing}) for the values $s=1, 0, -1$, respectively.  With the presence of the two frequency scales $\omega_1$ and $\omega_u\, (\ggg\omega_1)$, we have three different time range scales: the short time limit range $t\ll \omega_u^{-1}$, intermediate time range $ \omega_u^{-1}\ll t\ll \omega_1^{-1}$, and the long time limit range $t\gg \omega_1^{-1}$. Note that the high temperature limit corresponds to requiring $k_B T\gg \hbar\omega_1$ for the intermediate time range. We shall focus below on the intermediate and long time ranges, deriving analytical approximations to the induced phase and dephasing terms by expanding in frequency ratio parameter $\omega_1/\omega_u (\lll 1)$. The numerically evaluated sum of the two induced phase terms (\ref{analyticalkerr}) and (\ref{analyticalsecphase}) is plotted versus time in  Fig. \ref{inducedphasefig}, while the numerically evaluated  dephasing term integral expression given in the second line of Eq. (\ref{analyticaldephasing}) is plotted versus time in Fig. \ref{dephasingfig}. Both plots are normalized by their corresponding analytical approximations derived below in the $\omega_1t\rightarrow\infty$ limit, facilitating a check of the analytical approximations in the long time limit. The analytical approximations derived below for the net induced phase and dephasing terms are summarized in Table \ref{tabsummary}.

\begin{table*}
\begin{subtable}[h]{0.9\textwidth}
\begin{ruledtabular}
\centering
\begin{tabular}{ccc}
(a)&Net induced phase (intermediate time range) &Net induced phase (long time range)\\ \hline
$s=1$&$it(n+n'+1)(n-n') \frac{C \Omega^2 \omega_u }{\pi}$ & $it(n+n'+1)(n-n') \frac{C \Omega^2 \omega_u }{\pi}$ \\ 
$s=0$ & $it(n+n'+1)(n-n') \frac{C\Omega^2 }{\pi}\left[\ln(\omega_u t)-1\right]$
& $-it(n+n'+1)(n-n') \frac{C \Omega^2 }{\pi} \left[\ln \left( \omega_1/\omega_u\right)+\gamma\right]$\\ 
$s=-1$&$it^2(n+n'+1)(n-n') \frac{C\Omega^2}{4}$&$it(n+n'+1)(n-n')\frac{C \Omega^2 }{\pi\omega_1}$
\end{tabular}
\end{ruledtabular}
\end{subtable}

{\vskip 2mm}  
\begin{subtable}[h]{0.9\textwidth}
\begin{ruledtabular}
\begin{tabular}{ccc}
(b)&Dephasing term (intermediate time range)
&Dephasing term (long time range)\\ \hline
$s=1$&$-(n-n')^2C\Omega^2\left[\frac{1}{\pi} \ln\left(\beta\hbar\omega_u/2 \pi\right)+(\beta\hbar)^{-1}t \right]$&$-(n - n')^2\frac{2C \Omega^2}{\pi \beta \hbar \omega_1}$ \\ 
$s=0$&$-(n - n')^2\frac{C \Omega^2}{\pi \beta \hbar}  \left[\frac{3}{2} - {\gamma} -  \ln(\omega_1 t) \right]t^2$&$- (n - n')^2\frac{C \Omega^2}{\pi \beta \hbar \omega_1^2}$\\
$s=-1$&$ -(n - n')^2\frac{C \Omega^2}{\pi \omega_1 \beta \hbar}   t^2$&$-(n - n')^2\frac{2C \Omega^2}{3\pi \beta \hbar \omega_1^3}$
\end{tabular}
\end{ruledtabular}
\end{subtable}
\caption{\label{tabsummary}Leading order in $\omega_1/\omega_u$ expansion approximations to the net induced phase terms (a) and dephasing terms (b) in the intermediate time range ($\omega_u^{-1}\ll t\ll \omega_1^{-1}$) and long time range ($t\gg \omega_1^{-1}$) for ohmic ($s=1$) and subohmic ($s=0,\, -1$) bath spectral densities.}
\end{table*}

\subsection{Ohmic, $s=1$ environment case}
\label{ohmicsec}
We begin with the  ohmic case $s=1$, which corresponds to a 3D  acoustic environment medium. The first induced phase term (\ref{analyticalkerr}) is approximately $it(n+n'+1)(n-n') \frac{C \Omega^2 \omega_u }{\pi}$,
where we have expanded the incomplete Gamma function to leading order using the fact that $\omega_1/\omega_u \lll 1$. We see that this term diverges linearly with the upper frequency cut-off $\omega_u$.

In the intermediate time range ($\omega_u^{-1}\ll t\ll \omega_1^{-1}$), the second induced phase term (\ref{analyticalsecphase}) gives approximately
$-i(n+n'+1)(n-n') \frac{C\Omega^2}{2}$, while for the long time limit ($t\gg \omega_1^{-1}$) we obtain approximately $-i(n+n'+1)(n-n') \frac{C\Omega^2}{\pi} \frac{\cos(\omega_1 t)}{\omega_1 t}$; in both ranges, the second phase term is small compared to the above first phase term, as remarked previously.

The dephasing term (\ref{analyticaldephasing}) in the high temperature limit and intermediate time range becomes approximately
$-(n-n')^2 C\Omega^2\left[\frac{1}{\pi} \ln\left(\frac{\beta\hbar\omega_u}{2 \pi}\right)+(\beta\hbar)^{-1}t \right]$, with a leading linear dependence on time $t$. Note that in order to obtain the correct, logarithmically diverging term in $\omega_u$ appearing in the latter approximation, we instead used the exact solution to the dephasing term for $\omega_1=0$ derived in Ref. \cite{xu2020}.
In the long time limit ($t\gg \omega_1^{-1}$), the dephasing term  (\ref{analyticaldephasing}) becomes approximately $-(n - n')^2\frac{2C \Omega^2}{\pi \beta \hbar \omega_1}$.  Interestingly, this result is finite and independent of time, so that the final, reduced state $\rho_c$ of the cavity system mode will only be partially dephased in the Fock state basis. This is a consequence of the finite-sized volume of the acoustic environment medium, as signified by the non-zero fundamental frequency $\omega_1$ of the medium. We will see in the following that partial dephasing also occurs for the $s=0$ and $s=-1$ cases, again a consequence of the finite dimensions of the corresponding acoustic environments.

\begin{figure}
\begin{center}
\includegraphics[width=2.5in]{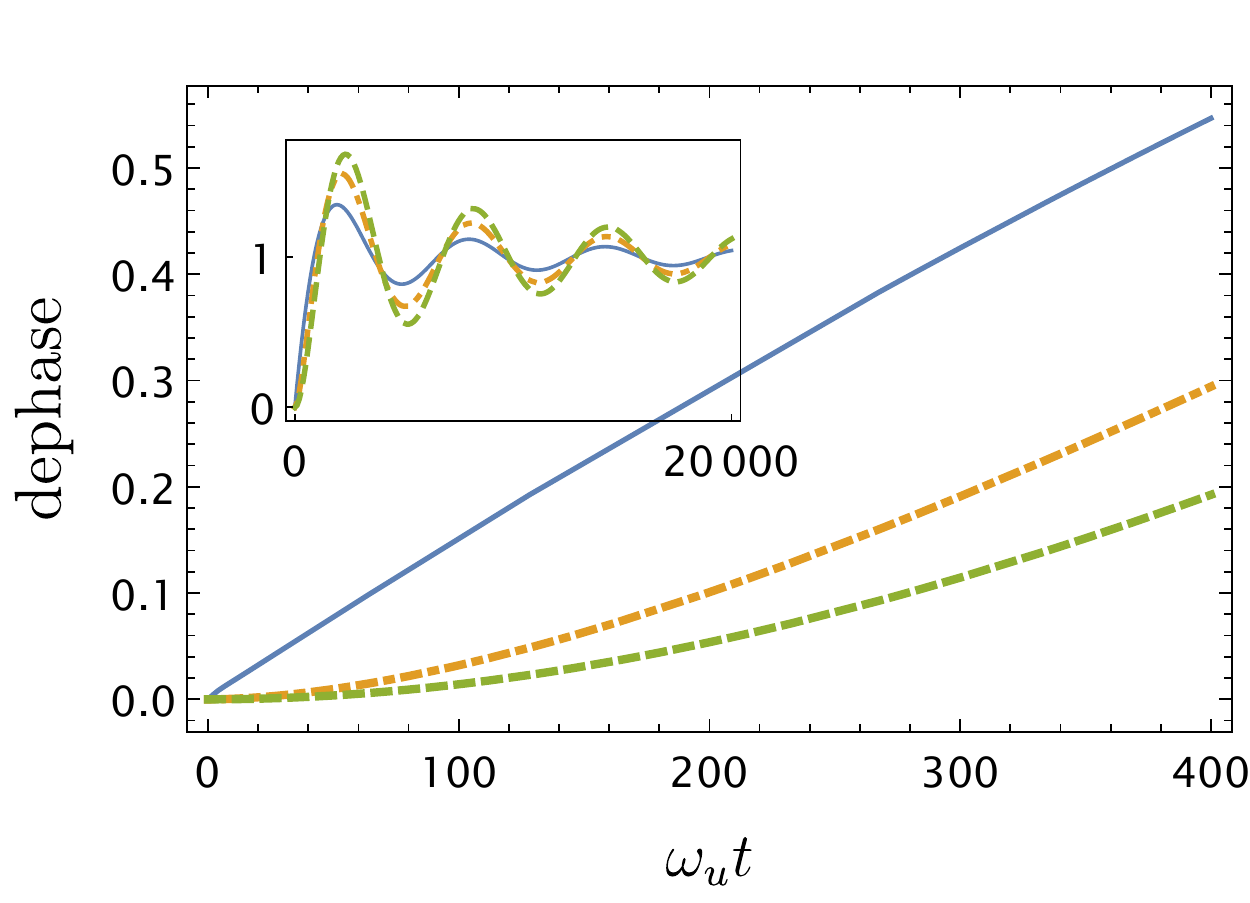} 
\includegraphics[width=0.8in]{legend.jpg}
\caption{\label{dephasingfig}The numerically evaluated, exact integral expression for the dephasing term given in Eq. (\ref{analyticaldephasing}) divided by its long time ($t\gg \omega_1^{-1}$) analytical expression as a function of the dimensionless time $\omega_u t$, with $\omega_1/\omega_u = 0.001$ and $\beta \hbar \omega_u = 10$. The inset gives the same normalized dephasing terms plotted over much longer timescales, indicating the expected approach to 1, hence validating the analytical approximation in the long time limit.}
\end{center}
\end{figure}

In Fig. \ref{dephasingfig}, the approach to the above-described, constant long time limit displays oscillatory behavior. This arises from the sub-leading contribution to the dephasing term, which takes the form $-(n - n')^2\frac{2C \Omega^2}{\pi \beta \hbar \omega_1}\times \frac{\sin(\omega_1 t)}{\omega_1 t}$. Oscillatory behavior also occurs for the $s=0$ and $s=-1$ cases as seen in Fig. \ref{dephasingfig},  arising  from similar sub-leading terms.

\subsection{Subohmic, $s=0$ environment case}
\label{subohmicsec}
For the subohmic $s=0$ case, which corresponds to a 2D acoustic environment medium,  the first induced phase term (\ref{analyticalkerr}) is approximately $-it(n+n'+1)(n-n') \frac{C \Omega^2 }{\pi} \left[\ln \left( \frac{\omega_1}{\omega_u}\right)+\gamma\right]$, to  leading order in an $\omega_1/\omega_u\, (\lll 1)$ expansion, where $\gamma\approx 0.5772\dots$ is the Euler-Mascheroni constant. Note that this phase term is both logarithmically UV ($\omega_u\rightarrow\infty$) and IR ($\omega_1\rightarrow 0$) divergent.  

For the intermediate time range ($\omega_u^{-1}\ll t\ll \omega_1^{-1}$), the second induced phase term (\ref{analyticalsecphase}) gives approximately $ it(n+n'+1)(n-n') \frac{C\Omega^2 }{\pi}\left[\ln (\omega_1 t) -1+\gamma\right]$. Combining with the above approximate expression for the first phase term, we obtain $it(n+n'+1)(n-n') \frac{C\Omega^2 }{\pi}\left[\ln(\omega_u t)-1\right]$, so that the net induced phase term is logarithmically divergent in the upper frequency cut-off $\omega_u$ for the intermediate time range.  In  the long time limit ($t\gg \omega_1^{-1}$) the phase term (\ref{analyticalsecphase}) approximates to $-i(n+n'+1)(n-n') \frac{C\Omega^2}{\pi}\frac{\cos \omega_1 t}{\omega_1^2 t}$. Again, we note that in the long time limit, this phase term becomes negligible compared with the first induced phase term. 

The dephasing term (\ref{analyticaldephasing}) in the high temperature limit and intermediate time range becomes approximately $-(n - n')^2\frac{C \Omega^2}{\pi \beta \hbar}  \left[\frac{3}{2} - {\gamma} -  \ln(\omega_1 t) \right]t^2$. In contrast to the corresponding $s=1$ dephasing term given in the previous subsection, the $s=0$ dephasing term is not UV divergent, but instead is IR divergent in the limit $\omega_1\rightarrow 0$. In the long time limit ($t\gg \omega_1^{-1}$), the dephasing term  (\ref{analyticaldephasing}) becomes approximately $- (n - n')^2\frac{C \Omega^2}{\pi \beta \hbar \omega_1^2}$.

\subsection{Subohmic, $s=-1$ environment case}
\label{subsubohmicsec}
For the subohmic $s=-1$ case, which corresponds to a 1D acoustic environment medium,  the first induced phase term (\ref{analyticalkerr}) is approximately $it(n+n'+1)(n-n')\frac{C \Omega^2 }{\pi\omega_1}$. In contrast to the corresponding $s=0$ phase term given in the previous subsection, this $s=-1$ phase term is IR divergent but not UV divergent.

For the intermediate time range ($\omega_u^{-1}\ll t\ll \omega_1^{-1}$), the second induced phase term (\ref{analyticalsecphase}) gives approximately $-it(n+n'+1)(n-n') \frac{C\Omega^2}{\pi\omega_1} \left[1-\frac{\pi}{4}\omega_1 t\right]$. Combining with the above approximate expression for the first phase term, we obtain for the net phase term: $it^2(n+n'+1)(n-n') \frac{C\Omega^2}{4}$, which is neither UV nor IR divergent. In  the  long  time  limit ($ t\gg \omega_1^{-1}$) the  phase  term  (8)  approximates  to  $-i(n+n'+1)(n-n') \frac{C\Omega^2 }{\pi}\frac{\cos \omega_1 t}{\omega_1^3 t}$, which becomes negligible compared with the first induced phase term.

The dephasing term (\ref{analyticaldephasing}) in the high temperature limit and intermediate time range becomes approximately $ -(n - n')^2\frac{C \Omega^2}{\pi \omega_1 \beta \hbar}   t^2$. Similarly to the corresponding $s=0$ dephasing term given in the previous subsection, the $s=-1$ dephasing term is IR divergent.  In the long time limit ($t\gg \omega_1^{-1}$), the dephasing term  (\ref{analyticaldephasing}) becomes approximately $-(n - n')^2\frac{2C \Omega^2}{3\pi \beta \hbar \omega_1^3}$.

\section{LC circuit--elastic strip model}
\label{1dcircuit}
\begin{figure}
\begin{center}
\includegraphics[width=3.5in]{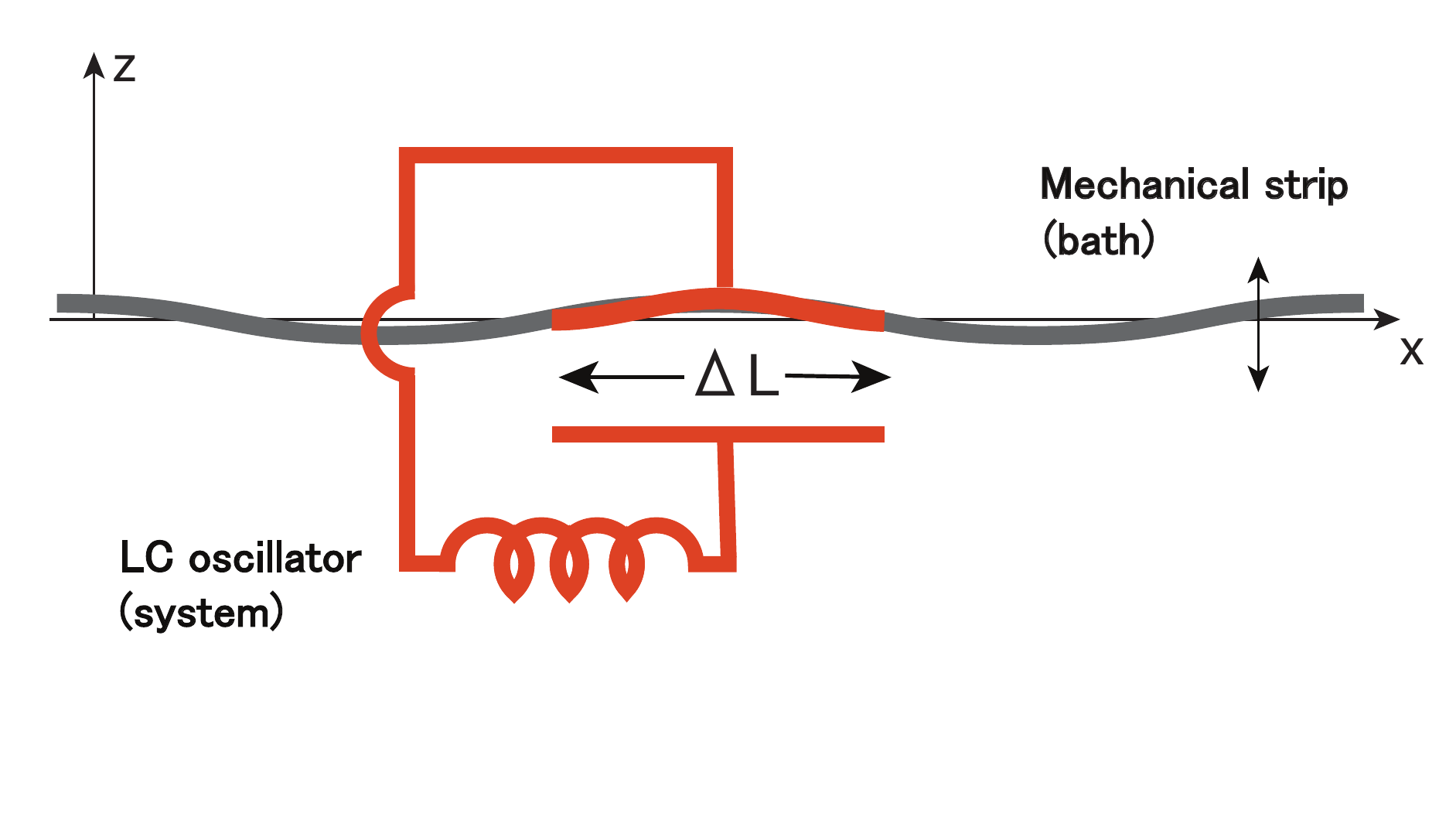} 
\caption{\label{fig1dmodel} Effectively 1D optomechanical scheme comprising an LC circuit oscillator (system) capacitively  coupled  to a long oscillating strip with (bath) via a metallized length $\Delta L$.}
\end{center}
\end{figure}
In this section we consider a model of a LC circuit capacitively coupled to a long mechanical strip (Fig.~\ref{fig1dmodel}). We show that this model system maps onto the subohmic $s=-1$ case considered in Sec. \ref{subsubohmicsec} (although with a different cut-off function and with some modifications to the integral approximation over the bath degrees of freedom). We will only consider dephasing, omitting the induced phase terms, i.e., cavity frequency renormalization and induced Kerr nonlinearity; the latter phase terms are orders of magnitude smaller than the bare LC circuit frequency phase term for the parameters considered later below in this section. We furthermore shall focus on dephasing during the intermediate time range only, where most of the dephasing occurs for the considered parameter values.

Referring to Fig.~\ref{fig1dmodel},  the lower conductor of the capacitor forming the LC circuit is assumed fixed, while the upper conductor is a flexing, metallized segment (length $\Delta L$) of a long elastic mechanical strip (length $L\ggg\Delta L$). The transverse width ($W$) and thickness ($T$) dimensions satisfy $T\ll W\lll L$.  The lower capacitor plate is assumed also to have length $\Delta L$ and the same width $W$ as the strip, with a small equilibrium  vacuum gap between upper and lower plates: $d\ll W,\,\Delta L$. The approximate mutual capacitance between the LC circuit and the undisplaced strip is approximately ${\mathsf{C}}_0 = \epsilon_0 W \Delta L/d$ and we denote the circuit inductance as ${\mathsf{L}}$. 

Neglecting motion in the transverse $y$ and longitudinal $x$ directions, we denote the flexing mechanical displacement field of the strip in the transverse $z$ direction by $u_z(x,t)$. For sufficiently large tensile forces $F$ applied at the clamped strip ends such that the elastic bending contribution can be neglected, the Lagrangian for the model, LC circuit-mechanical strip system in the resulting string-like limit is as follows:
\begin{align}
\mathcal{L} &= \frac{\rho_m W T}{2} \int_{0}^{L} dx \left(\frac{\partial u_z}{\partial t}\right)^2 \nn \\
&- \frac{F}{2} \int_0^{L} dx \left(\frac{\partial u_z}{\partial x} \right)^2 +\frac{1}{2}{\mathsf{C}}\left[u_z\right] \left(\frac{d \Phi}{dt} \right)^2 - \frac{\Phi^2}{2{\mathsf{L}}},
\label{originalLagforLCcircuit}
\end{align}
where ${\mathsf{C}}\left[u_z\right]$ denotes the mechanical displacement-dependent capacitance (with ${\mathsf{C}}\left[u_z=0\right]\equiv {\mathsf{C}}_0$ the equilibrium capacitance), $\Phi$ is the inductor flux coordinate, and $\rho_m$ is the mechanical strip mass density. 

Imposing fixed displacement field boundary conditions at the strip ends, $u_z(0) = u_z(L) = 0$, and solving for the free mechanical normal mode frequencies (see the appendix), we have
\begin{align}
\omega_i = \pi i \sqrt{\frac{F}{2 m L}},\, i=1,2,\dots,
\label{modestringeq}
\end{align}
with $m=\rho_m W T L/2$ the effective mass of the mechanical modes. Expanding the LC circuit frequency $\Omega=1/\sqrt{{\mathsf{L C}}}$
to first order in the displacement field $u_z$ and introducing mechanical mode and LC circuit creation/annihilation operators, the LC circuit-mechanical strip system Hamiltonian following from Lagrangian (\ref{originalLagforLCcircuit}) 
can be approximately mapped onto the optomechanical Hamiltonian (\ref{Hamiltonian}), with the coupling constant $\lambda_i$ taking the following form (see the appendix):
\begin{align}
\lambda_i = -\frac{1}{2 d} \left(\frac{\hbar}{2m\omega_i}\right)^{1/2} \sin\left(\frac{\pi i}{2} \right) \sinc  \left( \frac{\omega_i}{\omega_{u}}\right) ,\, i=1,2,\dots,
\label{expressionofgnfre1d}
\end{align}
where $ \sinc {x} := \sin x/x$ and   the upper cut-off frequency is
\begin{align}
\omega_u = \frac{2}{\Delta L} \sqrt{\frac{F L}{2m}}.
\label{omegaustringeq}
\end{align}

Comparing Eq. (\ref{omegaustringeq}) with the mode frequency expression (\ref{modestringeq}), we see that the upper cut-off frequency corresponds to the characteristic wavelength $\pi\Delta L$; in the limit where the mechanical mode wavelength becomes much smaller than the capacitor length $\Delta L$, the coupling between the cavity and mechanical strip spatially averages to zero, as expressed by the decaying sinc function appearing in Eq. (\ref{expressionofgnfre1d}).

With equally spaced, harmonic mode frequencies as given by Eq. (\ref{modestringeq}), we see from Eq. (\ref{outerprodeq}) that the dephasing term oscillates, completely vanishing at times $t=2\pi n/\omega_1, n=0,1,2,\dots$, where from Eq. (\ref{modestringeq}) the lower cut-off frequency is
\begin{align}
\omega_1 = \pi \sqrt{\frac{F}{2m L}}.
\label{lowerstripeq}
\end{align}
This periodic, full rephasing is to be contrasted with the non-zero, long time constant dephasing expressions obtained in Sec. \ref{dephasingsec}. The origin for this discrepancy is the breakdown of the integral approximation for the mode sums due to the strongly IR divergent nature of the latter appearing in Eq. (\ref{outerprodeq}) for the elastic strip model. An improved integral approximation for the mode sums can be obtained by employing the Euler-Maclaurin series formula to the desired order. In particular, utilizing Eq. (\ref{expressionofgnfre1d})  for $\lambda_i$ and the Euler-Maclaurin series approximation to first order for example, the integral of the bath spectral density approximation (\ref{generalspectraldensity}) in the large strip length $L$ limit is replaced by
\begin{align}
\pi\sum_i  \lambda_i^2& f(\omega_i) \approx C\int_{\omega_1}^{\infty} d\omega\, \omega^{-1} f(\omega)  \sinc ^2 \left( \frac{\omega}{\omega_u}\right) + C f(\omega_1),
\label{1dmodelspectraldensity}
\end{align}
where the coupling strength constant is
\begin{align}
C = \frac{\hbar}{8 d^2\sqrt{F\rho_m WT}} 
\label{couplingstripeq}
\end{align}
and we have approximated $\sinc(\omega_1/\omega_u) \approx 1$ since $\omega_1\ll \omega_u$.

Comparing the integral term in Eq. (\ref{1dmodelspectraldensity}) with Eq. (\ref{generalspectraldensity}), we see that the LC circuit-elastic strip (string) model corresponds to the $s=-1$ subohmic case, but with upper cut-off of the form $\sinc^2(\omega/\omega_u)$ instead of the previously considered exponential cut-off form $\exp(-\omega/\omega_u)$.
 Equation (\ref{1dmodelspectraldensity}) gives for the dephasing term in the intermediate time range ($\omega_u^{-1}\ll t\ll \omega_1^{-1}$): $-(n - n')^2\frac{2 C \Omega^2}{\pi \omega_1 \beta \hbar}   t^2$, approximately independent of the form of the upper cut-off. Note that the factor of 2 difference from the corresponding $s=-1$ dephasing expression given in Table \ref{tabsummary}b arises from the additional correction term in Eq. (\ref{1dmodelspectraldensity}); including higher order terms in the Euler-Maclaurin series approximation gives a factor closer to 2.5.    

From the $\omega_1^{-1}$ dependence of the analytical approximation to the $s=-1$ dephasing term (see Table \ref{tabsummary}b), it would seem that the dephasing rate can be made arbitrarily large by progressively increasing the strip length $L$. However, given that the optomechanical Hamiltonian approximation  (\ref{Hamiltonian}) results from expanding the LC circuit frequency to first order in the mechanical displacement field (i.e., weak coupling approximation), we necessarily require that mechanical induced fluctuations in the cavity frequency satisfy $\Delta\Omega\ll\Omega$. From Eqs. (\ref{Hamiltonian}) and (\ref{expressionofgnfre1d}), and assuming a thermal equilibrium state for the mechanical strip modes, the latter requirement gives (see the appendix for the derivation details): 
\begin{align}
\sum^{\infty}_{i=1} \frac{\hbar }{8m \omega_i d^2}\sin^2 \left( \frac{\pi i}{2}\right) \sinc^2\left(\frac{\omega_i}{\omega_u}\right)  \coth \left( \frac{\beta \hbar \omega_i}{2}\right) \ll 1,
\label{1dconstraint}
\end{align}
with $\omega_i$ and $\omega_u$ given by Eqs. (\ref{modestringeq}) and (\ref{omegaustringeq}) respectively.

In order to gain a sense of the dephasing rate magnitudes, we assume example parameter values similar to the silicon nitride vibrating string device of Ref. \cite{schilling16} (although allowing for much longer lengths $L$ than the actual $60~\mu{\mathrm{m}}$), and also assume typical superconducting microwave LC circuit parameters. In particular, we adopt the values $\rho_m=10^3~{\mathrm{kg}}/{\mathrm{m}}^3$, $F=10^{-5}~{\mathrm{N}}$,  $W=1~\mu{\mathrm{m}}$, $T=0.1~\mu{\mathrm{m}}$, and $L \gtrsim 1~{\mathrm{cm}}$. For the capacitor dimensions, we assume $\Delta L=10~\mu{\mathrm{m}}$ and $d=0.1~\mu{\mathrm{m}}$. The circuit mode frequency is assumed to be $\Omega/(2\pi)=5~{\mathrm{GHz}}$, and the acoustic bath temperature is taken to be $50~{\mathrm{mK}}$. With these assumed values, we have $\omega_i/(2\pi)=1.6 i\frac{{10\,\mathrm{cm}}}{L}~{\mathrm{kHz}}$ and $\omega_u/(2\pi)=10~{\mathrm{MHz}}$, giving $\omega_1/\omega_u=2\times 10^{-4}\frac{10\, {\mathrm{cm}}}{L}$. The dephasing term  then becomes approximately $-21 (n-n')^2 \frac{L}{10\,{\mathrm{cm}}} \frac{t^2}{\mu{\mathrm{s}}^2}$ in the intermediate time range $0.02\,\mu{\mathrm{s}}\ll t \ll 100 \frac{L}{10 {\mathrm{cm}}} \,\mu{\mathrm{s}}$. Thus we see that the phase interference between initial energy superposition states of the LC circuit mode is exponentially suppressed on timescales of microseconds for few centimeter long acoustic strip resonators. Rephasing occurs after a time $\approx 0.6 \frac{L}{10\,{\mathrm{cm}}}\,{\mathrm{msec}}$, neglecting other dephasing mechanisms.

Given that the LC circuit mode frequency satisfies $\Omega=500 \omega_u$, the cavity-mechanical oscillator bath interaction terms of the form $a^2 (b_i+b_i^{\dag})$ and $a^{\dag 2} (b_i+b_i^{\dag})$ may be neglected as discussed in the beginning of Sec. \ref{dephasingsec}. 
Furthermore, condition (\ref{1dconstraint}) on the strip length can be approximated as $L\ll 16\beta d^2 F\approx 2\times 10^6~{\mathrm{m}}$, which is orders of magnitude longer than in any conceivable circuit optomechanical device operating at cryogenic temperatures, and so the standard optomechanical interaction term in Eq. (\ref{Hamiltonian}) is well-justified. 

\section{Optical cavity--elastic membrane model}
\label{2dmembrane}
\begin{figure}
\begin{center}
\includegraphics[width=3.5in]{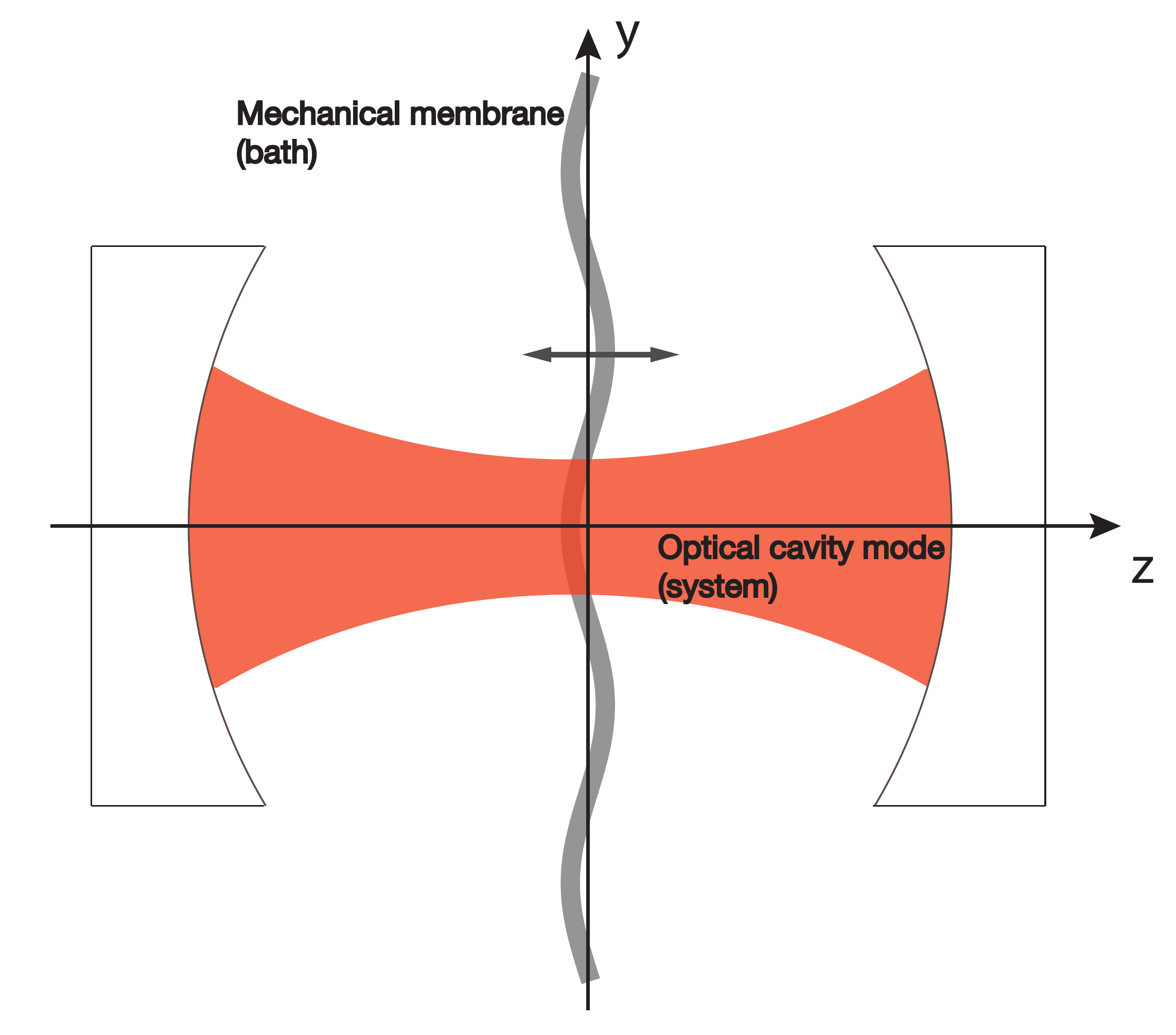} 
\caption{\label{2dcavitymem}Optomechanical scheme comprising a cavity light mode (system) trapped between oppositely facing mirrors interacting via light pressure with a thin dielectric membrane of large transverse extent and undergoing transverse flexural oscillations (bath).}
\label{optoexp}
\end{center}
\end{figure}

In this section we consider a model of a 3D optical cavity coupled to a large, square mechanical membrane  (Fig. \ref{2dcavitymem}) \cite{thompson2008}. We show that this model system maps onto the subohmic $s=0$ case considered in Sec. \ref{subsubohmicsec}. 
 As in the previous section, we will only consider dephasing, omitting the induced phase terms (i.e., cavity frequency renormalization and induced Kerr nonlinearity).

The cavity-membrane model system can be approximately described by the optomechanical Hamiltonian (\ref{Hamiltonian}) (see, e.g., Ref. [\onlinecite{underwoodthesis}]), with the mechanical normal mode frequencies of the vibrating membrane given by
\begin{align}
\omega_{i_x i_y} = \pi \sqrt{\frac{\mathcal{F}}{4 m} (i_x^2 + i_y^2)}, \, i_x, i_y=1,2,\dots,
\label{defimembranefrequency}
\end{align}
where $i_x,\, i_y$ are the mode labels marking the spatial dependencies of the modes in the transverse $x$ and $y$ coordinate dimensions of the membrane surface,  $\mathcal{F}$ is the tensile force per unit length applied at the clamped membrane edges and $m$ is the effective mass of the  mechanical modes:
\begin{align}
m = \rho_m L^2 T/4,
\label{2deffmass}
\end{align}
with the membrane having side dimension $L$ and thickness $T$; the tensile force is here assumed to be sufficiently large that the stretching potential energy dominates over the bending potential energy of the mechanical structure, hence defining the so-called membrane limit. 

Restricting to cavity  Gaussian beam modes, the cavity normal mode frequencies are approximately given by the following expression  \cite{brooker}:
\begin{align}
\Omega_{ \sigma} = \frac{\sigma \pi c}{l} + \frac{2c}{l} \tan^{-1}\left( \frac{l}{2f}\right),\, \sigma=1, 2,\dots,
\end{align}
where $l$ is the cavity length,  $f$ is a length parameter termed the ``Rayleigh range" that characterizes the mode beam profile, and $c$ is the speed of light in vacuum.

The optomechanical coupling between the Gaussian beam cavity modes (labeled by $\sigma$) and mechanical membrane modes (labeled by $i_x,\, i_y$) can be approximated as follows \cite{underwoodthesis}:
\begin{align}
\lambda_{\sigma, i_x i_y} &= (-1)^{\sigma}  \sqrt{\frac{\hbar }{2m \omega_{i_x i_y}}} \frac{(n^2 - 1)T\Omega_{ \sigma }}{l c}\sin \left(\frac{2\Omega_{\sigma }z_0}{c}\right)\nn \\
&\times\exp \left(-\frac{\omega_{i_x i_y}^2}{\omega_u^2} \right)
\sin\left(\frac{i_x \pi}{2}\right) \sin\left(\frac{i_y \pi}{2}\right),
\label{2domcoupleq}
\end{align}
where $z_0$ is the location of the membrane on the cavity's longitudinal axis, with  the membrane positioned such that its center coincides with the center of the cavity mode beam `waist' (i.e., the cavity midpoint with narrowest optical beam width defined as $w_{\sigma}=\sqrt{2f c/\Omega_{\sigma}}$), $n$ here denotes the membrane material optical index of refraction, and
\begin{align}
\omega_u = \sqrt{\frac{8\mathcal{F}}{\rho_m T w_{\sigma}^2}}
\label{2duppereq}
\end{align}
is the upper frequency cut-off. Expression (\ref{2domcoupleq}) assumes that the beam waist $w_{\sigma}$ is much smaller than the membrane side dimension $L$.

Comparing Eq. (\ref{2duppereq}) with the mechanical mode frequency expression (\ref{defimembranefrequency}), we see that the upper cut-off frequency corresponds to a mechanical mode wavelength comparable to the optical beam waist $w_{\sigma}$; in the limit where the mechanical mode wavelength becomes much smaller than the beam waist, the coupling between the cavity and mechanical membrane is exponentially suppressed as the square of the mode frequency.

The integral of the bath spectral density continuum approximation (\ref{generalspectraldensity}) gives
\begin{align}
\pi\sum_{i_x,i_y} \lambda_{\sigma,i_x i_y}^2 f(\omega_{i_x i_y}) &\approx  C \int_{\omega_1}^{\infty} d\omega  f(\omega) \exp\left(-\frac{2 \omega^2}{\omega_u^2}\right),
\label{2dmodelspectraldensity}
\end{align}
where from Eq. (\ref{defimembranefrequency}) the lower cut-off frequency is
\begin{align}
\omega_1 =  \pi \sqrt{\frac{\mathcal{F}}{2m}},
\end{align}
and the coupling strength constant is
\begin{align}
C = \frac{\hbar}{\mathcal{F}} \left[\frac{(n^2 - 1) \Omega_{\sigma} T \sin \left(\frac{2\Omega_{\sigma }z_0}{c}\right)}{2 l c}\right]^2.
\label{membranec}
\end{align}
Comparing the right hand sides of Eqs. (\ref{2dmodelspectraldensity}) and (\ref{generalspectraldensity}), we see that the optical cavity-elastic membrane model corresponds to the $s=0$ subohmic case, but with upper cut-off of the form $\exp(-2\omega^2/\omega^2_u)$ instead of the previously considered exponential cut-off form $\exp(-\omega/\omega_u)$.

Equation (\ref{2dmodelspectraldensity}) gives for the dephasing term in the intermediate time range ($\omega_u^{-1}\ll t\ll \omega_1^{-1}$): $-(n - n')^2\frac{1.3 C \Omega_{\sigma}^2}{\pi \beta \hbar}  \left[\frac{3}{2} - {\gamma} -  \ln(\omega_1 t) \right]t^2$, approximately independent of the form of the upper cut-off. 
The factor $1.3$ difference with the corresponding $s=0$ dephasing expression given in Table \ref{tabsummary}b accounts for the error in the continuous frequency integral approximation to the discrete sum over membrane modes given by Eq. (\ref{2dmodelspectraldensity}). This factor $1.3$ correction was simply determined by trial numerical fitting of the integral approximation over the intermediate time range, since there is no straightforward counterpart to the Euler-Maclaurin formula that gives the correction to the integral approximation of a double sum \cite{guo2021modified}.

In order to gain a sense of the dephasing rate magnitudes, we assume example parameter values similar to the silicon nitride vibrating membrane device of Ref. \cite{underwood2015measurement} (although allowing for much longer membrane side dimensions $L$ than the actual $1~{\mathrm{mm}}$). 
In particular, we adopt the values $n = 2$, $\rho_m=3.4\times 10^3~{\mathrm{kg}}/{\mathrm{m}}^3$, ${\mathcal{F}}=43~{\mathrm{N}}/{\mathrm{m}}$,  $T=50~{\mathrm{nm}}$, and $L \gtrsim 1~{\mathrm{cm}}$. For the optical mode, we assume a cavity length $l = 3.7~ \mathrm{cm}$ and infrared wavelength $\lambda_{\sigma}=1064~{\mathrm{nm}}$, corresponding to frequency $\Omega_{\sigma}/(2 \pi)=2.8\times 10^{14}~{\mathrm{Hz}}$ and beam waist $w_{\sigma}=90~\mu{\mathrm{m}}$, and suppose that the $z_0$ location of the membrane in the cavity is chosen such that the factor $|\sin(2\Omega_{\sigma} z_0/c)|=1$ in the coupling strength constant expression (\ref{membranec}). With these assumed values, we have $\omega_{i_x i_y}/(2\pi)=2.5 \sqrt{i_x^2+i_y^2}\frac{{10\,\mathrm{cm}}}{L}~{\mathrm{kHz}}$ and $\omega_u/(2\pi)=2.5~{\mathrm{MHz}}$, giving $\omega_1/\omega_u=1.4\times 10^{-3}\frac{10\, {\mathrm{cm}}}{L}$. 
The dephasing term then becomes approximately $-6 \times 10^{-6} (n-n')^2 \left[0.9-\ln\left(0.02 \frac{10\, {\mathrm{cm}}}{L}\frac{t}{\mu{\mathrm{s}}}\right)\right] \frac{T}{{\mathrm{K}}} \frac{t^2}{\mu{\mathrm{s}}^2}$ in the intermediate time range $0.06\,\mu{\mathrm{s}}\ll t \ll 45 \frac{L}{10 {\mathrm{cm}}} \,\mu{\mathrm{s}}$, where $\frac{T}{\mathrm{K}}$  refers to the membrane temperature expressed in Kelvin units. 
In the long time range $\omega_1^{-1}\ll t$ , the dephasing term oscillates strongly but does not completely vanish, in contrast to the strip case considered in Sec. \ref{1dcircuit}; due to the non-harmonic distribution of the membrane vibrational modes, complete rephasing does not occur.

From the just-derived expression for the dephasing term, we see that it scales approximately quadratically with the membrane edge length $L$ close to the  upper limit $\omega_1^{-1}$ of the intermediate time range. The resulting estimated dephasing term magnitudes for few centimeter scale-sized membranes are such that the contribution to dephasing of optical mode initial Fock state superposition states  due to the membrane environment is expected to be negligible compared to that of other sources, such as photon loss from the cavity. 

From the form of the coupling strength constant (\ref{membranec}), dephasing due to the membrane can also be increased somewhat by reducing the tensile force per unit length ${\mathcal{F}}$ applied to the membrane edges. However, the membrane approximation assumed in the present investigation  eventually breaks down as ${\mathcal{F}}$ is reduced; the bending potential energy contribution to the mechanical structure would need to be taken into account, with the structure behaving instead as a so-called plate having a qualitatively different flexural vibration mode spectrum.

Given that the cavity mode frequency satisfies $\Omega_{\sigma}=10^8\, \omega_u$, the cavity-mechanical oscillator bath interaction terms of the form $a^2 (b_i+b_i^{\dag})$ and $a^{\dag 2} (b_i+b_i^{\dag})$ may be neglected,  as discussed in the beginning of Sec. \ref{dephasingsec}. In contrast to the cavity-strip system considered in Sec.~\ref{1dcircuit}, the membrane induced fluctuations in the cavity mode frequency  remain constant with increasing membrane edge length $L$ (with the tensile force per unit length ${\mathcal{F}}$ kept fixed) and are negligible compared to the cavity mode frequency, so that there is no upper limit on the membrane edge length for the validity of the standard optomechanical interaction term in Eq. (\ref{Hamiltonian}).

\section{Conclusion}
\label{conclusionsec}

In the present work, we have investigated the quantum dynamics of optomechanical systems in the unusual situation where the mechanical subsystem comprises a dense spectrum of acoustic modes, functioning effectively as an environment for a single optical mode; in particular, the standard optomechanical interaction results in dephasing without dissipation of initial photon number superposition states of the optical mode. 

We found that the optical mode effective dynamics is qualitatively affected by the spatial dimension of the mechanical subsystem, with the dynamics for one dimensional mechanical environments (which can be realized for example as long elastic strings) exhibiting strong power law infrared divergences, two dimensional mechanical environments (such as large area elastic membranes) exhibiting weakly logarithmic  infrared and ultraviolet divergences, and three dimensional mechanical environments (such as large volume elastic solids) exhibiting strong power law ultraviolet divergences. The infrared divergences are regularized by accounting for the actual, finite size of the mechanical structures, characterized by the lowest mechanical mode frequency $\omega_1$. On the other hand, the ultraviolet divergences are regularized by the suppression of the optomechanical interaction on length scales smaller than the dimensions of the optomechanical interaction region, characterized by a given upper cut-off frequency $\omega_u (\gg \omega_1$).

We furthermore found that the cavity mode effective dynamics depends qualitatively on the time scales considered, with three different ranges delineated by the inverse frequencies $\omega_1^{-1}$ and $\omega_u^{-1}$. Dephasing predominantly occurs during the so-called `intermediate' range $\omega_u^{-1}\ll t\ll \omega_1^{-1}$, with a certain degree of rephasing occurring during the so-called long time range $\omega_1^{-1}\ll t$. 

Two possible realizations were considered in some detail, the first being a long elastic strip capacitively coupled to an LC circuit over a short segment of the strip, and an optical cavity mode coupled via light pressure to a large area elastic membrane. While the estimated dephasing rates resulting from these realizations are relatively small compared with photon loss rates from the cavities, they nevertheless afford useful model systems for clarifying our understanding of system-environment quantum dynamics for the unusual optomechanical type of interaction, where dephasing occurs without dissipation.  

We  also note that the models considered in this paper may  be interpreted as analogues for investigating various relativistic quantum information processes. For example, consider  two spatially separated cavities, each with a single mode--instead of just the one cavity mode--with both cavites coupled to the same mechanical structure. It would be interesting then to consider processes such as entanglement generation \cite{reznikEntanglementVacuum2003,martin-martinezSustainableEntanglementProduction2013,saltonAccelerationassistedEntanglementHarvesting2015a}, with the two cavities initially in a product of photon superposition states becoming entangled through their mutual interaction with the acoustic vacuum furnished by the extended mechanical structure. Because of the particular nature of the standard optomechanical interaction, such entanglement generation would occur in the absence of  real (or virtual) photon exchange, in contrast to the usually considered bilinear detector-field coupling.

\begin{acknowledgments}
We thank Sougato Bose for very helpful discussions. This work was supported by the NSF under grant no. PHY-2011382.
\end{acknowledgments}

\appendix
\section{LC circuit-elastic strip model}
\label{appendixLC}
\subsection{Derivation of the model Hamiltonian}
Starting from the Lagrangian in Eq.~(\ref{originalLagforLCcircuit}) and performing a Legendre transformation, the Hamiltonian for the model can be found as
\begin{align}
H =\int_0^{L}dx \left[ \frac{\pi_z(x,t)^2}{2\rho_m W T} +\frac{F}{2} \left(\frac{\partial u_z}{\partial x}\right)^2\right] + \frac{Q^2}{2\mathsf{C}[u_z]} + \frac{\Phi^2}{2\mathsf{L}}, \nn \\
\label{1dcircuitoriginalH}
\end{align}
where $Q$ and $\pi$ are the corresponding conjugate momenta for the flux and the displacement field:
\begin{subequations}
\begin{align}
Q &= \frac{\delta L}{\delta \dot{\Phi}}, \\
\pi_z &= \frac{\delta L}{\delta \dot{u}_z}.
\end{align}
\end{subequations}
Since we require that both ends of the strip are fixed with an applied tensile force $F$, the field $u_z$ then  satisfies the boundary condition: $u_z(0) = u_z(L) = 0$, and we can expand it in the normal mode basis as
\begin{align}
u_z(x,t) = \sum_{i=1}^{\infty} x_i(t) u_i(x),
\label{1dcircuitmechanicalexpand}
\end{align}
where $u_i(x) = \sin \left(\frac{\pi i x}{L}\right)$, $i=1,2,\dots$.
Substituting Eq.~(\ref{1dcircuitmechanicalexpand}) into Eq.~(\ref{1dcircuitoriginalH}), the strip Hamiltonian takes the
independent harmonic oscillator form:
\begin{align}
H=\sum_i \left(\frac{1}{2m}p_i^2 +\frac{1}{2}m \omega_i^2 x_i^2 \right) + \frac{Q^2}{2\mathsf{C}} + \frac{\Phi^2}{2\mathsf{L}} ,
\end{align}
where $p_i = m \frac{d x_i}{d t}$, $m$ is the mechanical mode effective strip mass:
\begin{align}
m = \frac{1}{2}\rho_m W T L,
\end{align}
and $\omega_i$ is the normal mode frequency:
\begin{align}
\omega_i = \frac{\pi i }{L} \sqrt{\frac{F}{\rho_m W T}}.
\end{align}

Quantization proceeds by promoting the coordinates $\Phi$, $x_i$ and their conjugate momenta into operators and imposing the usual commutation rules. Introducing the creation/annihilation operators defined by 
\begin{subequations}
\begin{align}
&Q = -i \left(\frac{\hbar}{2}\sqrt{\frac{\mathsf{C}}{\mathsf{L}}} \right)^{1/2} (a-a^\dag), \\
&\Phi = \left(\frac{\hbar}{2} \sqrt{\frac{\mathsf{L}}{\mathsf{C}}} \right)^{1/2}(a+ a^\dag),\\
&x_i = \left(\frac{\hbar}{2m\omega_i}\right)^{1/2}(b_i +b_i^\dag), \\
&p_i = -i \left(\frac{m\hbar \omega_i}{2} \right)^{1/2}(b_i- b_i^\dag) ,
\end{align}
\end{subequations}
the Hamiltonian simplifies to
\begin{align}
H = \hbar \Omega\left( a^\dag a + \frac{1}{2} \right) +\sum_n \hbar \omega_i \left( b_i^\dag b_i + \frac{1}{2} \right),
\end{align}
where $\Omega  = 1/\sqrt{\mathsf{LC}}$.

\subsection{Derivation of the coupling constant $\lambda_i$}
In order to obtain the optomechanical coupling between the LC circuit and the mechanical mode, we expand $\Omega$ to the first order in the normal mode displacement coordinates:
\begin{align}
\Omega &\approx \Omega_{0} + \sum_i \frac{\partial \Omega }{ \partial x_i} \bigg \rvert_{x_i =0} x_i  \nn \\
&= \frac{1}{\sqrt{\mathsf{L} \mathsf{C}_0}}  - \sum_i  \frac{\Omega_{0}}{2 \mathsf{C}_0} \frac{\partial \mathsf{C}}{\partial x_i} \bigg \rvert_{x_i = 0} \left(\frac{\hbar}{2m\omega_i}\right)^{1/2}(b_i + b_i^\dag)  \nn \\
&=\frac{1}{\sqrt{\mathsf{L} \mathsf{C}_0}}  + \sum_i \Omega_0 \lambda_i (b_n + b_n^\dag),
\label{definegn}
\end{align}
where we applied the chain rule in the second line of Eq.~(\ref{definegn}) and defined the coupling constant $\lambda_i$ in the last line. 

In order to determine the derivative of the capacitance, we shall first obtain an expression for the capacitance.
Assuming a positive charge $+Q$ placed on the upper conductor of the capacitor and a negative charge $-Q$ placed on the lower conductor, the electric field between the conductors can be found by solving the Laplace equation for the electric potential $\phi$:
\begin{align}
\frac{\partial^2 \phi}{ \partial z^2} = 0,
\end{align}
where we neglect the edge effects and approximate the electric field to be along the $z$ direction within the capacitor. 
With the lower strip at $z = -d$ and upper strip at $z = u_z(x)$, the boundary conditions for the electric potential are
\begin{subequations}
\begin{align}
&\phi(x, z = -d) = V_l,\\ 
&\phi(x, z = u_z(x)) = V_u,
\end{align}
\label{Voltagebound}
\end{subequations} 
where $V_l$, $V_u$ are the voltages on the lower and upper conductors. Since the displacement field $u_z$ is assumed to be much smaller than $d_0$, we can write the electric potential as a series expansion $\phi = \phi^{(0)} + \phi^{(1)} +...$. Substituting this series into the boundary conditions Eq.~(\ref{Voltagebound}), we have:
\begin{subequations}
\begin{align}
&\phi^{(0)} (x, -d) = V_l, \\ 
&\phi^{(0)}(x, 0) = V_u,
\end{align}
\end{subequations}
and 
\begin{subequations}
\begin{align}
&\phi^{(1)} (x, -d) = 0, \\ 
&\phi^{(1)}(x, 0) = -\frac{\partial \phi^{(0)}(x, z)}{\partial z} \bigg \rvert_{z = 0} u_z(x).
\end{align}
\end{subequations}
Solving the Laplace equation for $\phi^{(0)}$ and $\phi^{(1)}$, and taking the gradient, we obtain the electric field:
\begin{align}
\mathbf{E} &= -\nabla \left(\phi^{(0)} +\phi^{(1)} \right) \nn \\ 
&= -\frac{\Delta V}{d} \left(1 - \frac{u_z(x)}{d} \right) \hat{z},
\end{align}
where $\Delta V = V_u - V_l$.
In order to determine the relationship between the charge $Q$ and the voltage difference $\Delta V$, we apply  Gauss's law to a surface that just encloses the upper surface charge and we have:
\begin{align}
Q =& \frac{\epsilon_0 \Delta V W \Delta L}{d} - \frac{\epsilon_0 \Delta V W }{d^2} \int_{\frac{L - \Delta L}{2}}^{\frac{L +\Delta L}{2}} dx u_z(x) \nn \\ 
=& \mathsf{C}_0 \Delta V  -\frac{\mathsf{C}_0}{\Delta L d}\int_{\frac{L - \Delta L}{2}}^{\frac{L +\Delta L}{2}} dx u_z(x).
\label{expressionofQ}
\end{align}
With Eq.~(\ref{expressionofQ}), we have the expression for the capacitance:
\begin{align}
\mathsf{C} = \frac{Q}{\Delta V} = \mathsf{C}_0 -\frac{1}{\Delta L d}\int_{\frac{L - \Delta L}{2}}^{\frac{L +\Delta L}{2}} dx u_z(x).
\label{expressionofC}
\end{align}
Using the expansion for the displacement field Eq.~(\ref{1dcircuitmechanicalexpand}) and substituting Eq.~(\ref{expressionofC}) into Eq.~(\ref{definegn}), we find
\begin{align}
\lambda_i = - \frac{L} { \pi i d \Delta L }  \sinc \left( \frac{ \pi i \Delta L  }{2 L}\right) \sin \left(\frac{ \pi i }{2} \right)\left(\frac{\hbar}{2m\omega_i}\right)^{1/2},
\end{align}
where $ \sinc {x} := \sin x/x$. Expressing the coupling constant $\lambda_i$ in a frequency dependent form, we finally have the expression for $\lambda_i$ given by Eq.~(\ref{expressionofgnfre1d}):
\begin{align}
\lambda_i = -\frac{1}{2 d} \sinc \left( \frac{\omega_i}{\omega_u}\right) \sin\left(\frac{\pi i}{2} \right) \left(\frac{\hbar}{2m\omega_i}\right)^{1/2},
\end{align}
where the upper cut-off frequency is 
\begin{align}
\omega_u = \frac{2}{\Delta L} \sqrt{\frac{F}{\rho_m W T}}.  
\end{align} 

\subsection{Derivation of the strip length condition}
From Eq.~(\ref{definegn}), we have:
\begin{align}
\Omega &\approx  \Omega_0 + \sum_i \Omega_0 \lambda_i \left( \frac{2m\omega_n}{\hbar}\right)^{1/2} x_n.
\end{align}
Requiring that the variance of the capacitor frequency to be small compared with the square of its bare frequency $\Omega_{0}^2$, we have:
\begin{align}
\Bigg\langle \left(\sum_i \Omega_0 \lambda_i \left( \frac{2m\omega_i}{\hbar}\right)^{1/2} x_i\right)^2 \Bigg\rangle \ll \Omega_{0}^2.
\label{1dL0condition}
\end{align}
For a thermal harmonic oscillator with mass $m$ and frequency $\omega$, the variance for $x$ is:
\begin{align}
\langle x^2 \rangle =  \frac{\hbar}{2m\omega} \coth\left(\frac{\beta \hbar \omega}{2}\right),
\end{align}
so that Eq. (\ref{1dL0condition}) becomes
\begin{equation}
    \sum_i \lambda^2_i \coth\left(\frac{\beta \hbar \omega_i}{2}\right)\ll 1,
    \label{conditioneq}
\end{equation}
 where we use the fact that different mechanical modes are statistically independent.
Substituting the expression  (\ref{expressionofgnfre1d}) for $\lambda_i$ into Eq.~(\ref{1dL0condition}), we obtain condition (\ref{1dconstraint}):
\begin{align}
\sum_i \frac{\hbar }{8m \omega_i d^2} \sinc^2\left(\frac{\omega_i}{\omega_u}\right) \sin^2 \left( \frac{\pi i}{2}\right)^2 \coth \left( \frac{\beta \hbar \omega_i}{2}\right) \ll 1.
\end{align}

\bibliography{main}

\begin{thebibliography}{26}%
\makeatletter
\providecommand \@ifxundefined [1]{%
 \@ifx{#1\undefined}
}%
\providecommand \@ifnum [1]{%
 \ifnum #1\expandafter \@firstoftwo
 \else \expandafter \@secondoftwo
 \fi
}%
\providecommand \@ifx [1]{%
 \ifx #1\expandafter \@firstoftwo
 \else \expandafter \@secondoftwo
 \fi
}%
\providecommand \natexlab [1]{#1}%
\providecommand \enquote  [1]{``#1''}%
\providecommand \bibnamefont  [1]{#1}%
\providecommand \bibfnamefont [1]{#1}%
\providecommand \citenamefont [1]{#1}%
\providecommand \href@noop [0]{\@secondoftwo}%
\providecommand \href [0]{\begingroup \@sanitize@url \@href}%
\providecommand \@href[1]{\@@startlink{#1}\@@href}%
\providecommand \@@href[1]{\endgroup#1\@@endlink}%
\providecommand \@sanitize@url [0]{\catcode `\\12\catcode `\$12\catcode
  `\&12\catcode `\#12\catcode `\^12\catcode `\_12\catcode `\%12\relax}%
\providecommand \@@startlink[1]{}%
\providecommand \@@endlink[0]{}%
\providecommand \url  [0]{\begingroup\@sanitize@url \@url }%
\providecommand \@url [1]{\endgroup\@href {#1}{\urlprefix }}%
\providecommand \urlprefix  [0]{URL }%
\providecommand \Eprint [0]{\href }%
\providecommand \doibase [0]{http://dx.doi.org/}%
\providecommand \selectlanguage [0]{\@gobble}%
\providecommand \bibinfo  [0]{\@secondoftwo}%
\providecommand \bibfield  [0]{\@secondoftwo}%
\providecommand \translation [1]{[#1]}%
\providecommand \BibitemOpen [0]{}%
\providecommand \bibitemStop [0]{}%
\providecommand \bibitemNoStop [0]{.\EOS\space}%
\providecommand \EOS [0]{\spacefactor3000\relax}%
\providecommand \BibitemShut  [1]{\csname bibitem#1\endcsname}%
\let\auto@bib@innerbib\@empty
\bibitem [{\citenamefont {Aspelmeyer}\ \emph {et~al.}(2014)\citenamefont
  {Aspelmeyer}, \citenamefont {Kippenberg},\ and\ \citenamefont
  {Marquardt}}]{aspelmeyer2014}%
  \BibitemOpen
  \bibfield  {author} {\bibinfo {author} {\bibfnamefont {M.}~\bibnamefont
  {Aspelmeyer}}, \bibinfo {author} {\bibfnamefont {T.~J.}\ \bibnamefont
  {Kippenberg}}, \ and\ \bibinfo {author} {\bibfnamefont {F.}~\bibnamefont
  {Marquardt}},\ }\href@noop {} {\bibfield  {journal} {\bibinfo  {journal}
  {Reviews of Modern Physics}\ }\textbf {\bibinfo {volume} {86}},\ \bibinfo
  {pages} {1391} (\bibinfo {year} {2014})}\BibitemShut {NoStop}%
\bibitem [{\citenamefont {Bowen}\ and\ \citenamefont
  {Milburn}(2015)}]{bowen2015}%
  \BibitemOpen
  \bibfield  {author} {\bibinfo {author} {\bibfnamefont {W.~P.}\ \bibnamefont
  {Bowen}}\ and\ \bibinfo {author} {\bibfnamefont {G.~J.}\ \bibnamefont
  {Milburn}},\ }\href@noop {} {\emph {\bibinfo {title} {Quantum
  Optomechanics}}}\ (\bibinfo  {publisher} {CRC press},\ \bibinfo {year}
  {2015})\BibitemShut {NoStop}%
\bibitem [{\citenamefont {Renninger}\ \emph {et~al.}(2018)\citenamefont
  {Renninger}, \citenamefont {Kharel}, \citenamefont {Behunin},\ and\
  \citenamefont {Rakich}}]{renninger2018}%
  \BibitemOpen
  \bibfield  {author} {\bibinfo {author} {\bibfnamefont {W.}~\bibnamefont
  {Renninger}}, \bibinfo {author} {\bibfnamefont {P.}~\bibnamefont {Kharel}},
  \bibinfo {author} {\bibfnamefont {R.}~\bibnamefont {Behunin}}, \ and\
  \bibinfo {author} {\bibfnamefont {P.}~\bibnamefont {Rakich}},\ }\href@noop {}
  {\bibfield  {journal} {\bibinfo  {journal} {Nature Physics}\ }\textbf
  {\bibinfo {volume} {14}},\ \bibinfo {pages} {601} (\bibinfo {year}
  {2018})}\BibitemShut {NoStop}%
\bibitem [{\citenamefont {Gardiner}\ and\ \citenamefont
  {Zoller}(2004)}]{gardiner2004}%
  \BibitemOpen
  \bibfield  {author} {\bibinfo {author} {\bibfnamefont {C.}~\bibnamefont
  {Gardiner}}\ and\ \bibinfo {author} {\bibfnamefont {P.}~\bibnamefont
  {Zoller}},\ }\href@noop {} {\emph {\bibinfo {title} {Quantum Noise: a
  Handbook of Markovian and Non-Markovian Quantum Stochastic Methods with
  Applications to Quantum Optics}}},\ Vol.~\bibinfo {volume} {56}\ (\bibinfo
  {publisher} {Springer, Berlin},\ \bibinfo {year} {2004})\BibitemShut
  {NoStop}%
\bibitem [{\citenamefont {Bose}\ \emph {et~al.}(1997)\citenamefont {Bose},
  \citenamefont {Jacobs},\ and\ \citenamefont {Knight}}]{bose1997}%
  \BibitemOpen
  \bibfield  {author} {\bibinfo {author} {\bibfnamefont {S.}~\bibnamefont
  {Bose}}, \bibinfo {author} {\bibfnamefont {K.}~\bibnamefont {Jacobs}}, \ and\
  \bibinfo {author} {\bibfnamefont {P.}~\bibnamefont {Knight}},\ }\href@noop {}
  {\bibfield  {journal} {\bibinfo  {journal} {Physical Review A}\ }\textbf
  {\bibinfo {volume} {56}},\ \bibinfo {pages} {4175} (\bibinfo {year}
  {1997})}\BibitemShut {NoStop}%
\bibitem [{\citenamefont {Bose}\ \emph {et~al.}(1999)\citenamefont {Bose},
  \citenamefont {Jacobs},\ and\ \citenamefont {Knight}}]{bose1999scheme}%
  \BibitemOpen
  \bibfield  {author} {\bibinfo {author} {\bibfnamefont {S.}~\bibnamefont
  {Bose}}, \bibinfo {author} {\bibfnamefont {K.}~\bibnamefont {Jacobs}}, \ and\
  \bibinfo {author} {\bibfnamefont {P.~L.}\ \bibnamefont {Knight}},\
  }\href@noop {} {\bibfield  {journal} {\bibinfo  {journal} {Physical Review
  A}\ }\textbf {\bibinfo {volume} {59}},\ \bibinfo {pages} {3204} (\bibinfo
  {year} {1999})}\BibitemShut {NoStop}%
\bibitem [{\citenamefont {Blencowe}(2013)}]{blencowe2013}%
  \BibitemOpen
  \bibfield  {author} {\bibinfo {author} {\bibfnamefont {M.}~\bibnamefont
  {Blencowe}},\ }\href@noop {} {\bibfield  {journal} {\bibinfo  {journal}
  {Physical Review Letters}\ }\textbf {\bibinfo {volume} {111}},\ \bibinfo
  {pages} {021302} (\bibinfo {year} {2013})}\BibitemShut {NoStop}%
\bibitem [{\citenamefont {Xu}\ and\ \citenamefont {Blencowe}(2020)}]{xu2020}%
  \BibitemOpen
  \bibfield  {author} {\bibinfo {author} {\bibfnamefont {Q.}~\bibnamefont
  {Xu}}\ and\ \bibinfo {author} {\bibfnamefont {M.}~\bibnamefont {Blencowe}},\
  }\href@noop {} {\bibfield  {journal} {\bibinfo  {journal} {arXiv preprint
  arXiv:2005.02554}\ } (\bibinfo {year} {2020})}\BibitemShut {NoStop}%
\bibitem [{\citenamefont {Anastopoulos}\ and\ \citenamefont
  {Hu}(2013)}]{anastopoulos2013}%
  \BibitemOpen
  \bibfield  {author} {\bibinfo {author} {\bibfnamefont {C.}~\bibnamefont
  {Anastopoulos}}\ and\ \bibinfo {author} {\bibfnamefont {B.~L.}\ \bibnamefont
  {Hu}},\ }\href@noop {} {\bibfield  {journal} {\bibinfo  {journal} {Classical
  and Quantum Gravity}\ }\textbf {\bibinfo {volume} {30}},\ \bibinfo {pages}
  {165007} (\bibinfo {year} {2013})}\BibitemShut {NoStop}%
\bibitem [{\citenamefont {Michel}\ \emph {et~al.}(2015)\citenamefont {Michel},
  \citenamefont {Costamagna},\ and\ \citenamefont
  {Peeters}}]{michel2015theory}%
  \BibitemOpen
  \bibfield  {author} {\bibinfo {author} {\bibfnamefont {K.}~\bibnamefont
  {Michel}}, \bibinfo {author} {\bibfnamefont {S.}~\bibnamefont {Costamagna}},
  \ and\ \bibinfo {author} {\bibfnamefont {F.}~\bibnamefont {Peeters}},\
  }\href@noop {} {\bibfield  {journal} {\bibinfo  {journal} {Physica Status
  Solidi B}\ }\textbf {\bibinfo {volume} {252}},\ \bibinfo {pages} {2433}
  (\bibinfo {year} {2015})}\BibitemShut {NoStop}%
\bibitem [{\citenamefont {Clougherty}(2014)}]{clougherty2014quantum}%
  \BibitemOpen
  \bibfield  {author} {\bibinfo {author} {\bibfnamefont {D.~P.}\ \bibnamefont
  {Clougherty}},\ }\href@noop {} {\bibfield  {journal} {\bibinfo  {journal}
  {Physical Review B}\ }\textbf {\bibinfo {volume} {90}},\ \bibinfo {pages}
  {245412} (\bibinfo {year} {2014})}\BibitemShut {NoStop}%
\bibitem [{\citenamefont {Sengupta}\ \emph {et~al.}(2016)\citenamefont
  {Sengupta}, \citenamefont {Kotov},\ and\ \citenamefont
  {Clougherty}}]{sengupta2016infrared}%
  \BibitemOpen
  \bibfield  {author} {\bibinfo {author} {\bibfnamefont {S.}~\bibnamefont
  {Sengupta}}, \bibinfo {author} {\bibfnamefont {V.~N.}\ \bibnamefont {Kotov}},
  \ and\ \bibinfo {author} {\bibfnamefont {D.~P.}\ \bibnamefont {Clougherty}},\
  }\href@noop {} {\bibfield  {journal} {\bibinfo  {journal} {Physical Review
  B}\ }\textbf {\bibinfo {volume} {93}},\ \bibinfo {pages} {235437} (\bibinfo
  {year} {2016})}\BibitemShut {NoStop}%
\bibitem [{\citenamefont {Clougherty}\ and\ \citenamefont
  {Sengupta}(2017)}]{clougherty2017infrared}%
  \BibitemOpen
  \bibfield  {author} {\bibinfo {author} {\bibfnamefont {D.~P.}\ \bibnamefont
  {Clougherty}}\ and\ \bibinfo {author} {\bibfnamefont {S.}~\bibnamefont
  {Sengupta}},\ }\href@noop {} {\bibfield  {journal} {\bibinfo  {journal}
  {Physical Review A}\ }\textbf {\bibinfo {volume} {95}},\ \bibinfo {pages}
  {052110} (\bibinfo {year} {2017})}\BibitemShut {NoStop}%
\bibitem [{\citenamefont {Clougherty}(2017)}]{clougherty2017infraredtemp}%
  \BibitemOpen
  \bibfield  {author} {\bibinfo {author} {\bibfnamefont {D.~P.}\ \bibnamefont
  {Clougherty}},\ }\href@noop {} {\bibfield  {journal} {\bibinfo  {journal}
  {Physical Review B}\ }\textbf {\bibinfo {volume} {96}},\ \bibinfo {pages}
  {235404} (\bibinfo {year} {2017})}\BibitemShut {NoStop}%
\bibitem [{\citenamefont {Sengupta}\ and\ \citenamefont
  {Clougherty}(2017)}]{sengupta2017radiative}%
  \BibitemOpen
  \bibfield  {author} {\bibinfo {author} {\bibfnamefont {S.}~\bibnamefont
  {Sengupta}}\ and\ \bibinfo {author} {\bibfnamefont {D.~P.}\ \bibnamefont
  {Clougherty}},\ }\href@noop {} {\bibfield  {journal} {\bibinfo  {journal}
  {Physical Review B}\ }\textbf {\bibinfo {volume} {96}},\ \bibinfo {pages}
  {035419} (\bibinfo {year} {2017})}\BibitemShut {NoStop}%
\bibitem [{\citenamefont {Sengupta}(2019)}]{sengupta2019theory}%
  \BibitemOpen
  \bibfield  {author} {\bibinfo {author} {\bibfnamefont {S.}~\bibnamefont
  {Sengupta}},\ }\href@noop {} {\bibfield  {journal} {\bibinfo  {journal}
  {Physical Review B}\ }\textbf {\bibinfo {volume} {100}},\ \bibinfo {pages}
  {075429} (\bibinfo {year} {2019})}\BibitemShut {NoStop}%
\bibitem [{\citenamefont {Thompson}\ \emph {et~al.}(2008)\citenamefont
  {Thompson}, \citenamefont {Zwickl}, \citenamefont {Jayich}, \citenamefont
  {Marquardt}, \citenamefont {Girvin},\ and\ \citenamefont
  {Harris}}]{thompson2008}%
  \BibitemOpen
  \bibfield  {author} {\bibinfo {author} {\bibfnamefont {J.}~\bibnamefont
  {Thompson}}, \bibinfo {author} {\bibfnamefont {B.}~\bibnamefont {Zwickl}},
  \bibinfo {author} {\bibfnamefont {A.}~\bibnamefont {Jayich}}, \bibinfo
  {author} {\bibfnamefont {F.}~\bibnamefont {Marquardt}}, \bibinfo {author}
  {\bibfnamefont {S.}~\bibnamefont {Girvin}}, \ and\ \bibinfo {author}
  {\bibfnamefont {J.}~\bibnamefont {Harris}},\ }\href@noop {} {\bibfield
  {journal} {\bibinfo  {journal} {Nature}\ }\textbf {\bibinfo {volume} {452}},\
  \bibinfo {pages} {72} (\bibinfo {year} {2008})}\BibitemShut {NoStop}%
\bibitem [{\citenamefont {Leggett}\ \emph {et~al.}(1987)\citenamefont
  {Leggett}, \citenamefont {Chakravarty}, \citenamefont {Dorsey}, \citenamefont
  {Fisher}, \citenamefont {Garg},\ and\ \citenamefont {Zwerger}}]{leggett}%
  \BibitemOpen
  \bibfield  {author} {\bibinfo {author} {\bibfnamefont {A.~J.}\ \bibnamefont
  {Leggett}}, \bibinfo {author} {\bibfnamefont {S.}~\bibnamefont
  {Chakravarty}}, \bibinfo {author} {\bibfnamefont {A.~T.}\ \bibnamefont
  {Dorsey}}, \bibinfo {author} {\bibfnamefont {M.~P.}\ \bibnamefont {Fisher}},
  \bibinfo {author} {\bibfnamefont {A.}~\bibnamefont {Garg}}, \ and\ \bibinfo
  {author} {\bibfnamefont {W.}~\bibnamefont {Zwerger}},\ }\href@noop {}
  {\bibfield  {journal} {\bibinfo  {journal} {Reviews of Modern Physics}\
  }\textbf {\bibinfo {volume} {59}},\ \bibinfo {pages} {1} (\bibinfo {year}
  {1987})}\BibitemShut {NoStop}%
\bibitem [{\citenamefont {Schilling}\ \emph {et~al.}(2016)\citenamefont
  {Schilling}, \citenamefont {Sch{\"u}tz}, \citenamefont {Ghadimi},
  \citenamefont {Sudhir}, \citenamefont {Wilson},\ and\ \citenamefont
  {Kippenberg}}]{schilling16}%
  \BibitemOpen
  \bibfield  {author} {\bibinfo {author} {\bibfnamefont {R.}~\bibnamefont
  {Schilling}}, \bibinfo {author} {\bibfnamefont {H.}~\bibnamefont
  {Sch{\"u}tz}}, \bibinfo {author} {\bibfnamefont {A.}~\bibnamefont {Ghadimi}},
  \bibinfo {author} {\bibfnamefont {V.}~\bibnamefont {Sudhir}}, \bibinfo
  {author} {\bibfnamefont {D.~J.}\ \bibnamefont {Wilson}}, \ and\ \bibinfo
  {author} {\bibfnamefont {T.~J.}\ \bibnamefont {Kippenberg}},\ }\href@noop {}
  {\bibfield  {journal} {\bibinfo  {journal} {Physical Review Applied}\
  }\textbf {\bibinfo {volume} {5}},\ \bibinfo {pages} {054019} (\bibinfo {year}
  {2016})}\BibitemShut {NoStop}%
\bibitem [{\citenamefont {Underwood}(2016)}]{underwoodthesis}%
  \BibitemOpen
  \bibfield  {author} {\bibinfo {author} {\bibfnamefont {M.~J.}\ \bibnamefont
  {Underwood}},\ }\href@noop {} {\emph {\bibinfo {title} {Cryogenic
  optomechanics with a silicon nitride membrane}}}\ (\bibinfo  {publisher}
  {Ph.D. thesis, Yale University.},\ \bibinfo {year} {2016})\BibitemShut
  {NoStop}%
\bibitem [{\citenamefont {Brooker}(2003)}]{brooker}%
  \BibitemOpen
  \bibfield  {author} {\bibinfo {author} {\bibfnamefont {G.}~\bibnamefont
  {Brooker}},\ }\href@noop {} {\emph {\bibinfo {title} {Modern Classical
  Optics}}}\ (\bibinfo  {publisher} {Oxford University Press},\ \bibinfo
  {address} {Oxford},\ \bibinfo {year} {2003})\BibitemShut {NoStop}%
\bibitem [{\citenamefont {Guo}\ and\ \citenamefont
  {Liu}(2021)}]{guo2021modified}%
  \BibitemOpen
  \bibfield  {author} {\bibinfo {author} {\bibfnamefont {J.}~\bibnamefont
  {Guo}}\ and\ \bibinfo {author} {\bibfnamefont {Y.}~\bibnamefont {Liu}},\
  }\href@noop {} {\bibfield  {journal} {\bibinfo  {journal} {Communications in
  Theoretical Physics}\ }\textbf {\bibinfo {volume} {73}},\ \bibinfo {pages}
  {075002} (\bibinfo {year} {2021})}\BibitemShut {NoStop}%
\bibitem [{\citenamefont {Underwood}\ \emph {et~al.}(2015)\citenamefont
  {Underwood}, \citenamefont {Mason}, \citenamefont {Lee}, \citenamefont {Xu},
  \citenamefont {Jiang}, \citenamefont {Shkarin}, \citenamefont {B{\o}rkje},
  \citenamefont {Girvin},\ and\ \citenamefont
  {Harris}}]{underwood2015measurement}%
  \BibitemOpen
  \bibfield  {author} {\bibinfo {author} {\bibfnamefont {M.}~\bibnamefont
  {Underwood}}, \bibinfo {author} {\bibfnamefont {D.}~\bibnamefont {Mason}},
  \bibinfo {author} {\bibfnamefont {D.}~\bibnamefont {Lee}}, \bibinfo {author}
  {\bibfnamefont {H.}~\bibnamefont {Xu}}, \bibinfo {author} {\bibfnamefont
  {L.}~\bibnamefont {Jiang}}, \bibinfo {author} {\bibfnamefont
  {A.}~\bibnamefont {Shkarin}}, \bibinfo {author} {\bibfnamefont
  {K.}~\bibnamefont {B{\o}rkje}}, \bibinfo {author} {\bibfnamefont
  {S.}~\bibnamefont {Girvin}}, \ and\ \bibinfo {author} {\bibfnamefont
  {J.}~\bibnamefont {Harris}},\ }\href@noop {} {\bibfield  {journal} {\bibinfo
  {journal} {Physical Review A}\ }\textbf {\bibinfo {volume} {92}},\ \bibinfo
  {pages} {061801} (\bibinfo {year} {2015})}\BibitemShut {NoStop}%
\bibitem [{\citenamefont {Reznik}(2003)}]{reznikEntanglementVacuum2003}%
  \BibitemOpen
  \bibfield  {author} {\bibinfo {author} {\bibfnamefont {B.}~\bibnamefont
  {Reznik}},\ }\href {\doibase 10.1023/A:1022875910744} {\bibfield  {journal}
  {\bibinfo  {journal} {Foundations of Physics}\ }\textbf {\bibinfo {volume}
  {33}},\ \bibinfo {pages} {167} (\bibinfo {year} {2003})}\BibitemShut
  {NoStop}%
\bibitem [{\citenamefont {{Mart{\'i}n-Mart{\'i}nez}}\ \emph
  {et~al.}(2013)\citenamefont {{Mart{\'i}n-Mart{\'i}nez}}, \citenamefont
  {Brown}, \citenamefont {Donnelly},\ and\ \citenamefont
  {Kempf}}]{martin-martinezSustainableEntanglementProduction2013}%
  \BibitemOpen
  \bibfield  {author} {\bibinfo {author} {\bibfnamefont {E.}~\bibnamefont
  {{Mart{\'i}n-Mart{\'i}nez}}}, \bibinfo {author} {\bibfnamefont {E.~G.}\
  \bibnamefont {Brown}}, \bibinfo {author} {\bibfnamefont {W.}~\bibnamefont
  {Donnelly}}, \ and\ \bibinfo {author} {\bibfnamefont {A.}~\bibnamefont
  {Kempf}},\ }\href {\doibase 10.1103/PhysRevA.88.052310} {\bibfield  {journal}
  {\bibinfo  {journal} {Physical Review A}\ }\textbf {\bibinfo {volume} {88}},\
  \bibinfo {pages} {052310} (\bibinfo {year} {2013})}\BibitemShut {NoStop}%
\bibitem [{\citenamefont {Salton}\ \emph {et~al.}(2015)\citenamefont {Salton},
  \citenamefont {Mann},\ and\ \citenamefont
  {Menicucci}}]{saltonAccelerationassistedEntanglementHarvesting2015a}%
  \BibitemOpen
  \bibfield  {author} {\bibinfo {author} {\bibfnamefont {G.}~\bibnamefont
  {Salton}}, \bibinfo {author} {\bibfnamefont {R.~B.}\ \bibnamefont {Mann}}, \
  and\ \bibinfo {author} {\bibfnamefont {N.~C.}\ \bibnamefont {Menicucci}},\
  }\href {\doibase 10.1088/1367-2630/17/3/035001} {\bibfield  {journal}
  {\bibinfo  {journal} {New Journal of Physics}\ }\textbf {\bibinfo {volume}
  {17}},\ \bibinfo {pages} {035001} (\bibinfo {year} {2015})}\BibitemShut
  {NoStop}%
\end{thebibliography}%
\end{document}